# MATRIX-VECTOR REPRESENTATION
# OF VARIOUS SOLUTION CONCEPTS

Fuad Aleskerov, Andrey Subochev

*State University - Higher School of Economics*

A unified matrix-vector representation is developed of such solution concepts as the core, the uncovered, the uncaptured, the minimal weakly stable, the minimal undominated, the minimal dominant and the untrapped sets. We also propose several new versions of solution sets.

The work was partially supported by the Scientific Foundation of the State University – Higher School of Economics (grant № 08-04-0008), by Russian Foundation for Basic Research (joint Russian-Turkish research project, grant N 09-01-91224-CT_a) and by Decision Choice and Analysis Laboratory (DeCAN Lab) of the State University – Higher School of Economics.

*Fuad Aleskerov* – Department of Mathematics for Economics, University – Higher School of Economics (Moscow), alesk@hse.ru

*Andrey Subochev* – Department of Mathematics for Economics, University – Higher School of Economics (Moscow), asubochev@hse.ru

# 1. Introduction

In decision making theory solution concepts are of major significance. This stems from the fact that there is no single best solution for different decision making problems - each problem dictates its own reasonable answer.

In collective decision making the absence, in general case, of a maximal element in majority relation, i.e. nonexistence of an alternative more preferable for the majority of agents than any other alternative under binary comparisons, is called the Condorcet paradox. This very paradox led to proliferation of solution concepts over last 50 years of research in the area.

In this paper we develop a unified matrix-vector representation of such concepts as the core, the uncovered, the uncaptured, the minimal weakly stable, the minimal undominated, the minimal dominant and the untrapped sets, and propose several new versions of solution sets.

At the same time this representation determines a convenient algorithm for the calculation of solutions on majority relation.

The structure of the text is as follows. Basic definitions and notations are given in Section 2, where relations on the universal set of alternatives are considered in general. It is demonstrated how a relation and a subset of alternatives can be represented as a Boolean matrix and a Boolean vector, respectively. Also a vector-matrix representation for a set of maximal elements of an arbitrary relation is obtained in this Section.

Section 3 contains matrix-vector representations for the following solution sets: the Condorcet winner, the core, the five versions of the uncovered set (Fishburn, 1977; Miller, 1980), the uncaptured set (Duggan, 2007), the second version of the union of minimal weakly stable sets (Aleskerov, Kurbanov, 1999; Subochev, 2009), the union of minimal undominated sets (strong top-cycles) (Ward, 1961; Schwartz, 1970, 1972), the minimal dominant set (weak top-cycle) (Ward, 1961; Smith, 1973; Fishburn, 1977; Miller, 1977; Schwartz, 1977), the untrapped set (Duggan, 2007). These representations are obtained in general case, when ties are allowed.

In Section 4 new versions of some solution concepts are proposed: modifications of the five versions of the uncovered set and a new (third) version for a union of the minimal weakly stable sets. A criterion to determine whether an alternative belongs to the third version of the minimal weakly stable set is established. This criterion provides a connection between the union of the minimal weakly stable sets and the modified second version of the covering relation. The matrix-vector representations of the proposed solutions are also given.

Section 5 contains matrix-vector representations for the classes of k-stable alternatives and classes of k-stable sets introduced in Aleskerov, Subochev (2009) (see also Subochev (2009)). These classes are defined for tournaments, when there are no ties.

In Section 6 the results of the paper are summarized in the form of Theorem.

**2. Matrix-vector representation of sets and relations: basic definitions**

A finite set A of alternatives is given, $|A|=n>2$. Alternatives from A are denoted by a unique natural number i, $1 \leq i \leq n$, assigned to each of them. Throughout the paper plain lowercase letters without indices denote alternatives or numbers; plain capital letters without indices denote sets of alternatives.

A relation $\rho$ on A is a set of ordered pairs from A, $\rho \subseteq A \times A$. Throughout the paper Greek letters are used to denote relations. A relation $\rho$ is called symmetric if $\forall i, j \in A$ $(i, j) \in \rho \Rightarrow (j, i) \in \rho$. A relation is called asymmetric if it is irreflexive, i.e. $\forall i \in A$ $(i, i) \notin \rho$, and $\forall i, j \in A: i \neq j$ $(i, j) \in \rho \Rightarrow (j, i) \notin \rho$. Any relation $\rho$ can be unambiguously represented as a union of two relations, one of which is asymmetric and the other is symmetric. They are called asymmetric and symmetric parts of $\rho$ and denoted $\pi(\rho)$ and $\sigma(\rho)$, respectively: $\pi(\rho) \subseteq \rho$; $\sigma(\rho) \subseteq \rho$; $\pi(\rho) \cup \sigma(\rho) = \rho$; $(i, j) \in \rho$ & $(j, i) \notin \rho \Leftrightarrow (i, j) \in \pi(\rho)$; $(i, j) \in \rho$ & $(j, i) \in \rho \Leftrightarrow (i, j) \in \sigma(\rho)$. A relation $\rho$ is called complete if $\forall i, j \in A$ $i \neq j \Rightarrow i\rho j \vee j\rho i$. Evidently, if $\rho$ is a subrelation of $\omega$, $\rho \subseteq \omega \subseteq A \times A$, and if $\rho$ is complete, then $\omega$ is complete as well.

Any relation $\rho$ on a set A, $|A|=n$, can be uniquely represented by ($n \times n$) matrix. It is said that a matrix $\mathbf{R}=[r_{ij}]$ represents a relation $\rho$ if $r_{ij}=1 \Leftrightarrow (i, j) \in \rho$ and $r_{ij}=0 \Leftrightarrow (i, j) \notin \rho$.

Throughout the paper matrices are denoted by bold capital letters, matrix elements - by plain small letters with two indices. $\mathbf{E}$, $\mathbf{O}$ and $\mathbf{I}$ denote matrices $[e_{ij}]$, $[o_{ij}]$ and $[i_{ij}]$ such that $e_{ij}=1$ if $i=j$, 0 otherwise; $o_{ij}=0$ and $i_{ij}=1$ for any i, j∈A. Let ε denote the relation of identity, (i, j)∈ε ⇔ i=j. Evidently, the matrix **E** is a representation of the relation ε.

A set of alternatives B, B⊆A, can be represented by a characteristic n-component vector **b**=[$b_i$] defined in the following way: $b_i$=1 ⇔ i∈B and $b_i$=0 ⇔ i∉B. Let **e**(j) be a vector with only one (namely j'th) non-zero component and let this component be equal 1. Then **e**(j) will be a characteristic vector of a set containing only one alternative {j}. Let **a** denote a characteristic vector of the universal set A. By definition $a_i$=1 for any i, 1≤i≤n. Throughout the paper vectors are denoted by bold small letters, vector components - by plain small letters with one index.

Since we presume that all matrices and vectors are Boolean matrices and Boolean vectors, in all expressions, containing addition and/or multiplication of elements, these operations are understood as logical disjunction and conjunction, respectively. Addition and multiplication of matrices and vectors are defined and denoted in a standard way, for instance, $\mathbf{R}^i$ denotes a product of i matrices $\mathbf{R}^i = \underbrace{\mathbf{R} \cdot \mathbf{R} \cdot ... \cdot \mathbf{R}}_{i}$; $\mathbf{R}^{tr}$ denotes a transposed matrix: $\mathbf{Q}=\mathbf{R}^{tr}$ ⇔ $q_{ij}=r_{ji}$ for all i and j, 1≤i≤n, 1≤j≤n; diag(**R**) denotes a vector made of diagonal elements of **R**, i.e. **v**=diag(**R**) ⇔ $v_i=r_{ii}$ for all i, 1≤i≤n. $\overline{\mathbf{R}}$ and $\overline{\mathbf{v}}$ denote a matrix and a vector resulted after logical inversion of values of all elements in the corresponding matrix **R** and vector **v**, $\bar{r}_{ij}=0$ ⇔ $r_{ij}=1$. It is evident that transposition and logical inversion commute, $\overline{(\mathbf{R}^{tr})}=(\overline{\mathbf{R}})^{tr}$. Also $\mathbf{o}=\overline{\mathbf{a}}$, $o_i$=0 for any i, 1≤i≤n, is a characteristic vector for the empty set ∅. More generally, if **v** is a characteristic vector for a set V, V⊆A, then $\overline{\mathbf{v}}$ will be a characteristic vector for a set A\V. A characteristic vector of a union of sets is a sum of characteristic vectors of the sets united.

Let **P**(**R**) and **S**(**R**) denote matrices representing π(ρ) and σ(ρ), respectively. Lemma 1 contains main expressions for **R, P**(**R**) and **S**(**R**).

**Lemma 1.** $\mathbf{R}=\mathbf{P}(\mathbf{R})+\mathbf{S}(\mathbf{R})$; $\mathbf{S}(\mathbf{R})^{tr}=\mathbf{S}(\mathbf{R})$; $\mathbf{P}(\mathbf{R})=\overline{(\mathbf{R}^{tr}+\overline{\mathbf{R}})}$; $\mathbf{S}(\mathbf{R})=\overline{\overline{(\mathbf{R}^{tr}+\overline{\mathbf{R}})}}$. If ρ is complete, then $\mathbf{P}(\mathbf{R})+\mathbf{S}(\mathbf{R})+\mathbf{E}=\overline{\mathbf{P}^{tr}(\mathbf{R})}$, $\mathbf{P}(\mathbf{R})+\mathbf{P}^{tr}(\mathbf{R})+\mathbf{E}=\overline{\mathbf{S}(\mathbf{R})}$, $\mathbf{P}(\mathbf{R})+\mathbf{P}^{tr}(\mathbf{R})+\mathbf{S}(\mathbf{R})+\mathbf{E}=\mathbf{I}$.

Proof follows directly from definitions and the formula $\overline{a \vee b} = \overline{a} \wedge \overline{b}$.

An alternative i is called ρ-maximal in A iff $\forall j: j \neq i\; j\rho i \Rightarrow i\rho j$. Let MAX(ρ) denote a set of all ρ-maximal in A alternatives, i∈MAX(ρ) ⇔ i is ρ-maximal in A. If **R** is a matrix representing ρ, then i∉MAX(ρ) ⇔ ∃j: j≠i & $r_{ij}$=0 & $r_{ji}$=1. Let $\mathbf{Q} = \overline{\mathbf{R} + \overline{\mathbf{R}^{tr}}}$, then ∃j: j≠i & $r_{ij}$=0 & $r_{ji}$=1 ⇔ ∃j: $q_{ij}$=1. Then i∈MAX(ρ) ⇔ $q_{ij}$=0 for all j, 1≤j≤n. Let us multiply **Q** by **a**, **v**=**Q**·**a**. Since $a_i$=1 for all i, 1≤i≤n,

$v_i = \sum_{k=1}^{n} q_{ik} \cdot a_k = 0$ iff $q_{ij}$=0 for all j, 1≤j≤n. Then $v_i$=0 iff i∈MAX(ρ). Therefore $\overline{\mathbf{v}} = \overline{\mathbf{Q} \cdot \mathbf{a}} = \overline{\overline{(\mathbf{R} + \overline{\mathbf{R}^{tr}})} \cdot \mathbf{a}}$ =**max**(ρ) will be a characteristic vector for the set MAX(ρ).

This expression can be simplified if ρ is complete or asymmetric.

If ρ is complete then ∀j: j≠i & $r_{ij}$=0 ⇒ $r_{ji}$=1. Consequently, i∉MAX(ρ) ⇔ ∃j: j≠i & $r_{ij}$=0. Let $\mathbf{Q} = \overline{\mathbf{R} + \mathbf{E}}$, then ∃j: j≠i & $r_{ij}$=0 ⇔ ∃j: $q_{ij}$=1. Then i∈MAX(ρ) ⇔ $q_{ij}$=0 for all j, 1≤j≤n, therefore **max**(ρ)= $\overline{\mathbf{Q} \cdot \mathbf{a}} = \overline{\overline{(\mathbf{R} + \mathbf{E})} \cdot \mathbf{a}}$.

If ρ is asymmetric then ∀j: j≠i & $r_{ji}$=1 ⇒ $r_{ij}$=0, and $r_{ii}$=0 for all i∈A. Then i∉MAX(ρ) ⇔ ∃j: j≠i & $r_{ji}$=1. Let $\mathbf{Q} = \mathbf{R}^{tr}$, then ∃j: j≠i & $r_{ji}$=1 ⇔ ∃j: $q_{ij}$=1. Consequently, i∈MAX(ρ) ⇔ $q_{ij}$=0 for all j, 1≤j≤n, therefore **max**(ρ)= $\overline{\mathbf{Q} \cdot \mathbf{a}} = \overline{\mathbf{R}^{tr} \cdot \mathbf{a}}$.

Let us formulate this result as

**Lemma 2.** 1) If **R** is a matrix representing a relation ρ, then a characteristic vector **max**(ρ) for the set of its maximal elements MAX(ρ) is **max**(ρ)= $\overline{\overline{(\mathbf{R} + \overline{\mathbf{R}^{tr}})} \cdot \mathbf{a}}$ ; 2) if ρ is complete then **max**(ρ)= $\overline{\overline{(\mathbf{R} + \mathbf{E})} \cdot \mathbf{a}}$ ; 3) if ρ is asymmetric then **max**(ρ)= $\overline{\mathbf{R}^{tr} \cdot \mathbf{a}}$.

**Corollary.** MAX(ρ)=MAX(π(ρ)).

**Proof of the Corollary.** π(ρ) is asymmetric therefore by Lemma 2 **max**(α(ρ))= $\overline{\mathbf{P}^{tr}(\mathbf{R}) \cdot \mathbf{a}}$. By Lemma 1 $\mathbf{P}(\mathbf{R}) = \overline{(\mathbf{R}^{tr} + \overline{\mathbf{R}})}$. Since $\overline{(\mathbf{R}^{tr})} = (\overline{\mathbf{R}})^{tr}$, $\mathbf{P}^{tr}(\mathbf{R}) = (\overline{(\mathbf{R}^{tr} + \overline{\mathbf{R}})})^{tr} = \overline{(\mathbf{R}^{tr} + \overline{\mathbf{R}})^{tr}} = \overline{(\mathbf{R} + \overline{\mathbf{R}^{tr}})}$. Then **max**(π(ρ))= $\overline{\mathbf{P}^{tr}(\mathbf{R}) \cdot \mathbf{a}} = \overline{\overline{(\mathbf{R} + \overline{\mathbf{R}^{tr}})} \cdot \mathbf{a}}$ =**max**(ρ) ⇔ MAX(ρ)=MAX(π(ρ)).

An ordered pair (i, j) such that i$\rho$j is also called a $\rho$-step. A path from i to j is an ordered sequence of steps starting at i and ending at j, such that the second alternative in each step coincides with the first alternative of the next step. If all steps in a path belong to the same relation $\rho$, we call it a $\rho$-path. In other words a $\rho$-path is an ordered sequence of alternatives i, $j_1$, $j_2$, …, $j_{k-2}$, $j_{k-1}$, j, such that each alternative dominates the following one via $\rho$, i.e. i$\rho j_1$, $j_1 \rho j_2$, …, $j_{k-2}\rho j_{k-1}$, $j_{k-1}\rho j$. The number of steps in a path is called path's length. An alternative j is called reachable in k steps from i if there is a path of length k from i to j. A $\rho$-path from i to j is called a minimal $\rho$-path if i$\neq$j and there is no other $\rho$-path from i to j, which is shorter than the given one. By definition minimal $\rho$-paths are not cycles.

Let $\kappa(\rho)$ denote the transitive closure of $\rho$: (i, j)$\in\kappa(\rho)$ if j is reachable from i via $\rho$, i.e. if there is a $\rho$-path from i to j. We suppose that transitive closure is reflexive by definition, $\forall i\in A \Rightarrow$ (i, i)$\in\kappa(\rho)$. Let $\kappa_{(k)}(\rho)$ denote a k-transitive closure of $\rho$. A k-transitive closure is an abridged version of the transitive closure: (i, j)$\in\kappa_k(\rho) \Leftrightarrow$ i=j or j is reachable from i in no more than k steps via $\rho$, i.e. (i, j)$\in\kappa_k(\rho) \Leftrightarrow$ i=j or there is a $\rho$-path from i to j of length l: l$\leq$k. Evidently, if d is a maximum of lengths of all minimal $\rho$-paths in A, i.e. if d is a diameter of a digraph, which represents $\rho$, then a k-transitive closure of $\rho$ will be the transitive closure of $\rho$ iff k$\geq$d, $\kappa_k(\rho)=\kappa(\rho)$ $\Leftrightarrow$ k$\geq$d. The value d=d($\rho$) will be called a $\rho$-diameter of A.

### *Relations $\mu$, $\tau$ and $\upsilon$*

Now let us consider a framework of a collective decision making problem. A group of agents have to choose alternatives from the set A. The number of agents is greater than one. Each agent has preferences over alternatives from A. Majority relation is a binary relation $\mu$, $\mu\subset A\times A$, constructed as (i, j)$\in\mu$ if an alternative i is strongly preferred to an alternative j by majority, whichever defined, of all agents. If i$\mu$j then it is said that i dominates j and j is dominated by i. By assumption majority is defined so that $\mu$ is asymmetric. If neither (i, j)$\in\mu$, nor (j, i)$\in\mu$ holds, then (i, j) is called a tie. A set of ties $\tau$ is a symmetric binary relation on A: $\tau\subseteq A\times A$, (i, j)$\in\tau \Rightarrow$ (j, i)$\in\tau$.

By definition both μ and τ are irreflexive, i.e. (i, i)∉μ, (i, i)∉τ for all i∈A. Let υ denote a relation, which is a union of μ, τ and ε, υ=μ∪τ∪ε. It follows from definitions of μ, τ and ε that υ is complete, reflexive and μ=π(υ), τ∪ε=σ(υ) hold.

A relation μ is called a tournament, if it is complete. Thus μ is a tournament when corresponding τ is empty, τ=∅, which is equivalent to μ∪ε=υ.

Let **M**=[$m_{ij}$] denote a matrix representing μ: $m_{ij}$=1 if an alternative, which corresponds to a row i, dominates an alternative, which corresponds to a column j, and $m_{ij}$=0 if i is dominated by j or ties it. **T**=[$t_{ij}$] and **U**=[$u_{ij}$] will denote the matrices representing τ and υ, respectively. Evidently, **M**=**P**(**U**) and **T**+**E**=**S**(**U**), therefore according to Lemma 1 the following expressions hold:
$$\mathbf{U}=\mathbf{M}+\mathbf{T}+\mathbf{E}=\overline{\mathbf{M}^{tr}}, \quad \mathbf{M}+\mathbf{M}^{tr}+\mathbf{E}=\overline{\mathbf{T}}, \quad \mathbf{M}+\mathbf{M}^{tr}+\mathbf{T}+\mathbf{E}=\mathbf{I}.$$

To illustrate this let us consider the following example: A={1, 2, 3, 4, 5, 6}, μ={(1, 2), (2, 3), (3, 1), (4, 1), (4, 2), (4, 5), (5, 6), (6, 2), (6, 4)} (see Figure 1). Then

$$\mathbf{M}=\begin{pmatrix} 0 & 1 & 0 & 0 & 0 & 0 \\ 0 & 0 & 1 & 0 & 0 & 0 \\ 1 & 0 & 0 & 0 & 0 & 0 \\ 1 & 1 & 0 & 0 & 1 & 0 \\ 0 & 0 & 0 & 0 & 0 & 1 \\ 0 & 1 & 0 & 1 & 0 & 0 \end{pmatrix}, \mathbf{T}=\begin{pmatrix} 0 & 0 & 0 & 0 & 1 & 1 \\ 0 & 0 & 0 & 0 & 1 & 0 \\ 0 & 0 & 0 & 1 & 1 & 1 \\ 0 & 0 & 1 & 0 & 0 & 0 \\ 1 & 1 & 1 & 0 & 0 & 0 \\ 1 & 0 & 1 & 0 & 0 & 0 \end{pmatrix}, \mathbf{U}=\begin{pmatrix} 1 & 1 & 0 & 0 & 1 & 1 \\ 0 & 1 & 1 & 0 & 1 & 0 \\ 1 & 0 & 1 & 1 & 1 & 1 \\ 1 & 1 & 1 & 1 & 1 & 0 \\ 1 & 1 & 1 & 0 & 1 & 1 \\ 1 & 1 & 1 & 1 & 0 & 1 \end{pmatrix}$$

$$\mathbf{M}^{tr}=\begin{pmatrix} 0 & 0 & 1 & 1 & 0 & 0 \\ 1 & 0 & 0 & 1 & 0 & 1 \\ 0 & 1 & 0 & 0 & 0 & 0 \\ 0 & 0 & 0 & 0 & 0 & 1 \\ 0 & 0 & 0 & 1 & 0 & 0 \\ 0 & 0 & 0 & 0 & 1 & 0 \end{pmatrix}, \overline{\mathbf{M}}=\begin{pmatrix} 1 & 0 & 1 & 1 & 1 & 1 \\ 1 & 1 & 0 & 1 & 1 & 1 \\ 0 & 1 & 1 & 1 & 1 & 1 \\ 0 & 0 & 1 & 1 & 0 & 1 \\ 1 & 1 & 1 & 1 & 1 & 0 \\ 1 & 0 & 1 & 0 & 1 & 1 \end{pmatrix}, \overline{\mathbf{T}}=\begin{pmatrix} 1 & 1 & 1 & 1 & 0 & 0 \\ 1 & 1 & 1 & 1 & 0 & 1 \\ 1 & 1 & 1 & 0 & 0 & 0 \\ 1 & 1 & 0 & 1 & 1 & 1 \\ 0 & 0 & 0 & 1 & 1 & 1 \\ 0 & 1 & 0 & 1 & 1 & 1 \end{pmatrix}$$

A lower contour set of an alternative i is a set L(i) of all alternatives dominated by i, L(i)={j∈A: iμj}. Correspondingly, an upper contour set of an alternative i is a set D(i) of all alternatives dominating i, D(i)={j∈A: jμi}. A horizon of i is a set H(i) of all alternatives j, for which (i, j) is a tie, H(i)={j∈A: iτj}. Obviously, L(i)∪D(i)∪H(i)∪{i}=A. Let **l**(i), **d**(i) and **h**(i) denote characteristic vectors of L(i), D(i) and H(i), respectively. These vectors can be calculated by the following formulae.

(1)    $\mathbf{l}(i)=\mathbf{M}^{tr}\cdot\mathbf{e}(i)$, $\mathbf{d}(i)=\mathbf{M}\cdot\mathbf{e}(i)$, $\mathbf{h}(i)=\mathbf{T}\cdot\mathbf{e}(i)$

The proof is obvious. It should be also noted that the expression $L(i)\cup D(i)\cup H(i)\cup\{i\}=A$, which must hold for any i, can be represented as $\mathbf{l}(i)+\mathbf{d}(i)+\mathbf{h}(i)+\mathbf{e}(i)=\mathbf{a}$. The latter directly follows from the formulae (1) and $\mathbf{M}^{tr}+\mathbf{M}+\mathbf{T}+\mathbf{E}=\mathbf{I}$.

In the example above for the alternative 2 we obtain

$$\mathbf{l}(2)=\mathbf{M}^{tr}\cdot\mathbf{e}(2)=\begin{pmatrix}0&0&1&1&0&0\\1&0&0&1&0&1\\0&1&0&0&0&0\\0&0&0&0&0&1\\0&0&0&1&0&0\\0&0&0&0&1&0\end{pmatrix}\cdot\begin{pmatrix}0\\1\\0\\0\\0\\0\end{pmatrix}=\begin{pmatrix}0\\0\\1\\0\\0\\0\end{pmatrix}=\mathbf{e}(3)\Leftrightarrow L(2)=\{3\}$$

$$\mathbf{d}(2)=\mathbf{M}\cdot\mathbf{e}(2)=\begin{pmatrix}0&1&0&0&0&0\\0&0&1&0&0&0\\1&0&0&0&0&0\\1&1&0&0&1&0\\0&0&0&0&0&1\\0&1&0&1&0&0\end{pmatrix}\cdot\begin{pmatrix}0\\1\\0\\0\\0\\0\end{pmatrix}=\begin{pmatrix}1\\0\\0\\1\\0\\1\end{pmatrix}=\mathbf{e}(1)+\mathbf{e}(4)+\mathbf{e}(6)\Leftrightarrow D(2)=\{1,4,6\}$$

$$\mathbf{h}(2)=\mathbf{T}\cdot\mathbf{e}(2)=\begin{pmatrix}0&0&0&0&1&1\\0&0&0&0&1&0\\0&0&0&1&1&1\\0&0&1&0&0&0\\1&1&1&0&0&0\\1&0&1&0&0&0\end{pmatrix}\cdot\begin{pmatrix}0\\1\\0\\0\\0\\0\end{pmatrix}=\begin{pmatrix}0\\0\\0\\0\\1\\0\end{pmatrix}=\mathbf{e}(4)\Leftrightarrow H(2)=\{5\}$$

**3. Representations for various solution concepts in general case**

*The Condorcet winner and the core*

The core Cr is defined as a set of all undominated alternatives in A, $i\in Cr \Leftrightarrow D(i)=\emptyset$. That is $Cr=MAX(\mu)$. Since $\mu=\pi(\upsilon)$, then by Lemma 2 and its Corollary $Cr=MAX(\mu)=MAX(\upsilon)$ and

$$\mathbf{cr}=\mathbf{max}(\mu)=\mathbf{max}(\upsilon)=\overline{\overline{\mathbf{M}^{tr}\cdot\mathbf{a}}}=\overline{\overline{\mathbf{U}\cdot\mathbf{a}}}=\overline{\overline{(\mathbf{M}+\mathbf{T}+\mathbf{E})\cdot\mathbf{a}}}.$$

In the example considered above we obtain

$$\overline{\mathbf{M}+\mathbf{T}+\mathbf{E}}=\mathbf{M}^{tr}=\begin{pmatrix} 0 & 0 & 1 & 1 & 0 & 0 \\ 1 & 0 & 0 & 1 & 0 & 1 \\ 0 & 1 & 0 & 0 & 0 & 0 \\ 0 & 0 & 0 & 0 & 0 & 1 \\ 0 & 0 & 0 & 1 & 0 & 0 \\ 0 & 0 & 0 & 0 & 1 & 0 \end{pmatrix},$$

$$\mathbf{cr}=\overline{\overline{(\mathbf{M}+\mathbf{T}+\mathbf{E})}\cdot \mathbf{a}}=\overline{\begin{pmatrix} 0 & 0 & 1 & 1 & 0 & 0 \\ 1 & 0 & 0 & 1 & 0 & 1 \\ 0 & 1 & 0 & 0 & 0 & 0 \\ 0 & 0 & 0 & 0 & 0 & 1 \\ 0 & 0 & 0 & 1 & 0 & 0 \\ 0 & 0 & 0 & 0 & 1 & 0 \end{pmatrix}\cdot\begin{pmatrix}1\\1\\1\\1\\1\\1\end{pmatrix}}=\overline{\begin{pmatrix}\overline{1}\\\overline{1}\\\overline{1}\\\overline{1}\\\overline{1}\\\overline{1}\end{pmatrix}}=\begin{pmatrix}0\\0\\0\\0\\0\\0\end{pmatrix}=\mathbf{o}\Leftrightarrow \text{Cr}=\varnothing.$$

That is in our example the core is empty.

If a matrix **R** is represented as a sum of some matrices **X** and **Y**, i.e. **R=X+Y**, then $r_{ij}=1 \Leftrightarrow x_{ij}=1 \vee y_{ij}=1$. Let **R=M+E**. If $r_{ij}=1$ for any j: j≠i, then i is a Condorcet winner, i.e. an alternative dominating any other alternative. Therefore $\mathbf{cw}=\overline{\overline{(\mathbf{M}+\mathbf{E})}\cdot \mathbf{a}}$ is a characteristic vector of a set of Condorcet winners.

In our example

$$\mathbf{cw}=\overline{\overline{(\mathbf{M}+\mathbf{E})}\cdot \mathbf{a}}=\overline{\begin{pmatrix} 0 & 0 & 1 & 1 & 1 & 1 \\ 1 & 0 & 0 & 1 & 1 & 1 \\ 0 & 1 & 0 & 1 & 1 & 1 \\ 0 & 0 & 1 & 0 & 0 & 1 \\ 1 & 1 & 1 & 1 & 0 & 0 \\ 1 & 0 & 1 & 0 & 1 & 0 \end{pmatrix}\cdot\begin{pmatrix}1\\1\\1\\1\\1\\1\end{pmatrix}}=\overline{\begin{pmatrix}\overline{1}\\\overline{1}\\\overline{1}\\\overline{1}\\\overline{1}\\\overline{1}\end{pmatrix}}=\begin{pmatrix}0\\0\\0\\0\\0\\0\end{pmatrix}=\mathbf{o}.$$

Consequently, {CW}=∅, i.e. a Condorcet winner does not exist.

*The uncovered set*

Calculating the uncovered set in a tournament Banks (1986) considered a product of matrices **R=M·M** and pointed out that an element $r_{ij}=\sum_{k=1}^{n} m_{ik}\cdot m_{kj}$ is not equal to zero iff there is at list one alternative k such that i dominates k and k dominates j. That is $r_{ij} \neq 0$ iff there is a two-step μ-path from i to j: iμk & kμj. Since we presume that all vectors and matrices are Boolean ones, $r_{ij}=1$ iff there is a two-step μ-path from i to j and $r_{ij}=0$ iff there is no such path. Respectively, if **R**=**M**·**T** then $r_{ij}=1$ iff there is a two-step path from i to j, where the first step is a μ-step iμk, and the second step is a τ-step kτj. In other words, $r_{ij}=1 \Leftrightarrow \exists k, k \in A$, such that iμk & kτj, otherwise $r_{ij}=0$. Analogously if **R**=**T**·**M** then $r_{ij}=1 \Leftrightarrow \exists k, k \in A$: iτk & kμj, and if **R**=**T**·**T** then $r_{ij}=1 \Leftrightarrow \exists k, k \in A$: iτk & kτj, otherwise $r_{ij}=0$.

Five versions of the covering relation have been given in order to define uncovered alternatives in general case. Let us denote them as $\alpha^I$, $\alpha^{II}$, $\alpha^{III}$, $\alpha^{IV}$ and $\alpha^V$, respectively.

Version 1: i covers j, $(i, j) \in \alpha^I \Leftrightarrow$ iμj & $L(j) \subseteq L(i) \cup H(i)$, then i is uncovered $\Leftrightarrow \forall j$: jμi $\Rightarrow \exists k$: iμk & kμj (Duggan, 2007).

Version 2: i covers j, $(i, j) \in \alpha^{II} \Leftrightarrow$ iμj & $L(j) \subseteq L(i)$, then i is uncovered $\Leftrightarrow \forall j$: jμi $\Rightarrow \exists k$: (iμk & kμj)∨(iμk & kτj) (Miller, 1980).

Version 3: i covers j, $(i, j) \in \alpha^{III} \Leftrightarrow$ iμj & $D(i) \subseteq D(j)$, then i is uncovered $\Leftrightarrow \forall j$: jμi $\Rightarrow \exists k$: (iμk & kμj)∨(iτk & kμj) (Fishburn, 1977; Miller, 1980).

Version 4: i covers j, $(i, j) \in \alpha^{IV} \Leftrightarrow$ iμj & $L(j) \subseteq L(i)$ & $D(i) \subseteq D(j)$, then i is uncovered $\Leftrightarrow \forall j$: jμi $\Rightarrow \exists k$: (iμk & kμj)∨(iμk & kτj)∨(iτk & kμj) (Miller, 1980; McKelvey, 1986).

Version 5: i covers j, $(i, j) \in \alpha^V \Leftrightarrow$ iμj & $H(j) \cup L(j) \subseteq L(i)$, then i is uncovered $\Leftrightarrow \forall j$: jμi $\Rightarrow \exists k$: (iμk & kμj)∨(iμk & kτj)∨(iτk & kμj)∨(iτk & kτj) (Duggan, 2007).

An uncovered alternative is an alternative, which is not covered by any alternative from A. The uncovered set UC is comprised of all uncovered alternatives from A. $UC^I$, $UC^{II}$, $UC^{III}$, $UC^{IV}$ and $UC^V$ denote uncovered sets derived from the first to fifth definitions of the covering relation, respectively. Since all versions of the covering relation are asymmetric, uncovered alternatives and only they are maximal elements of α, that is $UC^I=MAX(\alpha^I)$, $UC^{II}=MAX(\alpha^{II})$, $UC^{III}=MAX(\alpha^{III})$, $UC^{IV}=MAX(\alpha^{IV})$, and $UC^V=MAX(\alpha^V)$.

Let us construct matrices representing the relations $\alpha^I$, $\alpha^{II}$, $\alpha^{III}$, $\alpha^{IV}$ and $\alpha^V$. Let $\mathbf{Q}=\mathbf{M}^2+\mathbf{M}+\mathbf{T}+\mathbf{E}$. If $q_{ij}=1$ then either $i=j$, or $i\tau j$, or $i\mu j$, or $\exists k$: $i\mu k$ & $k\mu j$ hold. Consequently, if $q_{ij}=1$, then i is not covered by j according to the first version of the covering relation, $(j, i)\notin\alpha^I$. If there is an alternative j, such that $q_{ij}=0$, then neither $i=j$, nor $i\tau j$, nor $i\mu j$ holds, hence $j\mu i$. Also $\sum_{k=1}^{n} m_{ik} \cdot m_{kj} = 0$ $\Rightarrow$ ($m_{kj}=1 \Rightarrow m_{ik}=0$) $\Rightarrow$ ($\forall k$: $k\mu j \Rightarrow (k\mu i \vee k\tau i)$). Therefore, if $q_{ij}=0$, then i is covered by j according to the first version of the covering relation, $(j, i)\in\alpha^I$. Then ($q_{ij}=1 \Rightarrow (j, i)\notin\alpha^I$ and $q_{ij}=0 \Rightarrow (j, i)\in\alpha^I$) $\Leftrightarrow \overline{\mathbf{Q}}^{tr} = \overline{(\mathbf{M}\cdot\mathbf{M}+\mathbf{M}+\mathbf{T}+\mathbf{E})}^{tr} = \mathbf{R}$ is a matrix representation of $\alpha^I$. Similar considerations lead us to the following conclusion: $\overline{(\mathbf{M}\cdot\mathbf{T}+\mathbf{M}\cdot\mathbf{M}+\mathbf{M}+\mathbf{T}+\mathbf{E})}^{tr}$, $\overline{(\mathbf{T}\cdot\mathbf{M}+\mathbf{M}\cdot\mathbf{M}+\mathbf{M}+\mathbf{T}+\mathbf{E})}^{tr}$, $\overline{(\mathbf{T}\cdot\mathbf{M}+\mathbf{M}\cdot\mathbf{T}+\mathbf{M}\cdot\mathbf{M}+\mathbf{M}+\mathbf{T}+\mathbf{E})}^{tr}$ and $\overline{(\mathbf{T}\cdot\mathbf{T}+\mathbf{T}\cdot\mathbf{M}+\mathbf{M}\cdot\mathbf{T}+\mathbf{M}\cdot\mathbf{M}+\mathbf{M}+\mathbf{T}+\mathbf{E})}^{tr}$ are matrices representing relations $\alpha^{II}$, $\alpha^{III}$, $\alpha^{IV}$ and $\alpha^V$, respectively.

Since all versions of the covering relation are asymmetric, by Lemma 2 we obtain the following formulae for characteristic vectors $\mathbf{uc}^I$, $\mathbf{uc}^{II}$, $\mathbf{uc}^{III}$, $\mathbf{uc}^{IV}$ and $\mathbf{uc}^V$ of the uncovered sets $UC^I$, $UC^{II}$, $UC^{III}$, $UC^{IV}$ and $UC^V$:

$$\mathbf{uc}=\mathbf{max}(\alpha)= \overline{\mathbf{R}^{tr} \cdot \mathbf{a}} = \overline{\overline{\mathbf{Q}} \cdot \mathbf{a}},$$

$$\mathbf{uc}^I=\mathbf{max}(\alpha^I)= \overline{\overline{(\mathbf{M}\cdot\mathbf{M}+\mathbf{M}+\mathbf{T}+\mathbf{E})} \cdot \mathbf{a}},$$

$$\mathbf{uc}^{II}=\mathbf{max}(\alpha^{II})= \overline{\overline{(\mathbf{M}\cdot\mathbf{T}+\mathbf{M}\cdot\mathbf{M}+\mathbf{M}+\mathbf{T}+\mathbf{E})} \cdot \mathbf{a}},$$

$$\mathbf{uc}^{III}=\mathbf{max}(\alpha^{III})= \overline{\overline{(\mathbf{T}\cdot\mathbf{M}+\mathbf{M}\cdot\mathbf{M}+\mathbf{M}+\mathbf{T}+\mathbf{E})} \cdot \mathbf{a}},$$

$$\mathbf{uc}^{IV}=\mathbf{max}(\alpha^{IV})= \overline{\overline{(\mathbf{T}\cdot\mathbf{M}+\mathbf{M}\cdot\mathbf{T}+\mathbf{M}\cdot\mathbf{M}+\mathbf{M}+\mathbf{T}+\mathbf{E})} \cdot \mathbf{a}} \text{ and}$$

$$\mathbf{uc}^V=\mathbf{max}(\alpha^V)= \overline{\overline{(\mathbf{T}\cdot\mathbf{T}+\mathbf{T}\cdot\mathbf{M}+\mathbf{M}\cdot\mathbf{T}+\mathbf{M}\cdot\mathbf{M}+\mathbf{M}+\mathbf{T}+\mathbf{E})} \cdot \mathbf{a}}.$$

For the example considered above we obtain

$$\mathbf{M}\cdot\mathbf{M}=\begin{pmatrix}0&1&0&0&0&0\\0&0&1&0&0&0\\1&0&0&0&0&0\\1&1&0&0&1&0\\0&0&0&0&0&1\\0&1&0&1&0&0\end{pmatrix}\cdot\begin{pmatrix}0&1&0&0&0&0\\0&0&1&0&0&0\\1&0&0&0&0&0\\1&1&0&0&1&0\\0&0&0&0&0&1\\0&1&0&1&0&0\end{pmatrix}=\begin{pmatrix}0&0&1&0&0&0\\1&0&0&0&0&0\\0&1&0&0&0&0\\0&1&1&0&0&1\\0&1&0&1&0&0\\1&1&1&0&1&0\end{pmatrix},$$

$$\mathbf{M}\cdot\mathbf{T}=\begin{pmatrix}0&1&0&0&0&0\\0&0&1&0&0&0\\1&0&0&0&0&0\\1&1&0&0&1&0\\0&0&0&0&0&1\\0&1&0&1&0&0\end{pmatrix}\cdot\begin{pmatrix}0&0&0&0&1&1\\0&0&0&0&1&0\\0&0&0&1&1&1\\0&0&1&0&0&0\\1&1&1&0&0&0\\1&0&1&0&0&0\end{pmatrix}=\begin{pmatrix}0&0&0&0&1&0\\0&0&0&1&1&1\\0&0&0&0&1&1\\1&1&1&0&1&1\\1&0&1&0&0&0\\0&0&1&0&1&0\end{pmatrix},$$

$$\mathbf{T}\cdot\mathbf{M}=\begin{pmatrix}0&0&0&0&1&1\\0&0&0&0&1&0\\0&0&0&1&1&1\\0&0&1&0&0&0\\1&1&1&0&0&0\\1&0&1&0&0&0\end{pmatrix}\cdot\begin{pmatrix}0&1&0&0&0&0\\0&0&1&0&0&0\\1&0&0&0&0&0\\1&1&0&0&1&0\\0&0&0&0&0&1\\0&1&0&1&0&0\end{pmatrix}=\begin{pmatrix}0&1&0&1&0&1\\0&0&0&0&0&1\\1&1&0&1&1&1\\1&0&0&0&0&0\\1&1&1&0&0&0\\1&1&0&0&0&0\end{pmatrix},$$

$$\mathbf{M}^2+\mathbf{M}+\mathbf{T}+\mathbf{E}=\begin{pmatrix}0&0&1&0&0&0\\1&0&0&0&0&0\\0&1&0&0&0&0\\0&1&1&0&0&1\\0&1&0&1&0&0\\1&1&1&0&1&0\end{pmatrix}+\begin{pmatrix}1&1&0&0&1&1\\0&1&1&0&1&0\\1&0&1&1&1&1\\1&1&1&1&1&0\\1&1&1&0&1&1\\1&1&1&1&0&1\end{pmatrix}=\begin{pmatrix}1&1&1&0&1&1\\1&1&1&0&1&0\\1&1&1&1&1&1\\1&1&1&1&1&1\\1&1&1&1&1&1\\1&1&1&1&1&1\end{pmatrix}$$

$$\mathbf{M}\cdot\mathbf{T}+\mathbf{M}^2+\mathbf{M}+\mathbf{T}+\mathbf{E}=\begin{pmatrix}0&0&0&0&1&0\\0&0&0&1&1&1\\0&0&0&0&1&1\\1&1&1&0&1&1\\1&0&1&0&0&0\\0&0&1&0&1&0\end{pmatrix}+\begin{pmatrix}1&1&1&0&1&1\\1&1&1&0&1&0\\1&1&1&1&1&1\\1&1&1&1&1&1\\1&1&1&1&1&1\\1&1&1&1&1&1\end{pmatrix}=\begin{pmatrix}1&1&1&0&1&1\\1&1&1&1&1&1\\1&1&1&1&1&1\\1&1&1&1&1&1\\1&1&1&1&1&1\\1&1&1&1&1&1\end{pmatrix},$$

$$\mathbf{T}\cdot\mathbf{M}+\mathbf{M}^2+\mathbf{M}+\mathbf{T}+\mathbf{E}=\begin{pmatrix}0&1&0&1&0&1\\0&0&0&0&0&1\\1&1&0&1&1&1\\1&0&0&0&0&0\\1&1&1&0&0&0\\1&1&0&0&0&0\end{pmatrix}+\begin{pmatrix}1&1&1&0&1&1\\1&1&1&0&1&0\\1&1&1&1&1&1\\1&1&1&1&1&1\\1&1&1&1&1&1\\1&1&1&1&1&1\end{pmatrix}=\begin{pmatrix}1&1&1&1&1&1\\1&1&1&0&1&1\\1&1&1&1&1&1\\1&1&1&1&1&1\\1&1&1&1&1&1\\1&1&1&1&1&1\end{pmatrix},$$

$$\mathbf{T \cdot M + M \cdot T + M^2 + M + T + E} = \begin{pmatrix} 0 & 1 & 0 & 1 & 0 & 1 \\ 0 & 0 & 0 & 0 & 0 & 1 \\ 1 & 1 & 0 & 1 & 1 & 1 \\ 1 & 0 & 0 & 0 & 0 & 0 \\ 1 & 1 & 1 & 0 & 0 & 0 \\ 1 & 1 & 0 & 0 & 0 & 0 \end{pmatrix} + \begin{pmatrix} 1 & 1 & 1 & 0 & 1 & 1 \\ 1 & 1 & 1 & 1 & 1 & 1 \\ 1 & 1 & 1 & 1 & 1 & 1 \\ 1 & 1 & 1 & 1 & 1 & 1 \\ 1 & 1 & 1 & 1 & 1 & 1 \\ 1 & 1 & 1 & 1 & 1 & 1 \end{pmatrix} = \begin{pmatrix} 1 & 1 & 1 & 1 & 1 & 1 \\ 1 & 1 & 1 & 1 & 1 & 1 \\ 1 & 1 & 1 & 1 & 1 & 1 \\ 1 & 1 & 1 & 1 & 1 & 1 \\ 1 & 1 & 1 & 1 & 1 & 1 \\ 1 & 1 & 1 & 1 & 1 & 1 \end{pmatrix} = \mathbf{I},$$

$$\mathbf{uc^I} = \overline{\overline{(\mathbf{M \cdot M + M + T + E}) \cdot \mathbf{a}}} =$$

$$= \overline{\overline{\begin{pmatrix} 0 & 0 & 0 & 1 & 0 & 0 \\ 0 & 0 & 0 & 1 & 0 & 1 \\ 0 & 0 & 0 & 0 & 0 & 0 \\ 0 & 0 & 0 & 0 & 0 & 0 \\ 0 & 0 & 0 & 0 & 0 & 0 \\ 0 & 0 & 0 & 0 & 0 & 0 \end{pmatrix} \cdot \begin{pmatrix} 1 \\ 1 \\ 1 \\ 1 \\ 1 \\ 1 \end{pmatrix}}} = \overline{\begin{pmatrix} \overline{1} \\ \overline{1} \\ \overline{0} \\ \overline{0} \\ \overline{0} \\ \overline{0} \end{pmatrix}} = \begin{pmatrix} 0 \\ 0 \\ 1 \\ 1 \\ 1 \\ 1 \end{pmatrix} \Leftrightarrow UC^I = \{3, 4, 5, 6\},$$

$$\mathbf{uc^{II}} = \overline{\overline{(\mathbf{M \cdot T + M \cdot M + M + T + E}) \cdot \mathbf{a}}} =$$

$$= \overline{\overline{\begin{pmatrix} 0 & 0 & 0 & 1 & 0 & 0 \\ 0 & 0 & 0 & 0 & 0 & 0 \\ 0 & 0 & 0 & 0 & 0 & 0 \\ 0 & 0 & 0 & 0 & 0 & 0 \\ 0 & 0 & 0 & 0 & 0 & 0 \\ 0 & 0 & 0 & 0 & 0 & 0 \end{pmatrix} \cdot \begin{pmatrix} 1 \\ 1 \\ 1 \\ 1 \\ 1 \\ 1 \end{pmatrix}}} = \overline{\begin{pmatrix} \overline{1} \\ \overline{0} \\ \overline{0} \\ \overline{0} \\ \overline{0} \\ \overline{0} \end{pmatrix}} = \begin{pmatrix} 0 \\ 1 \\ 1 \\ 1 \\ 1 \\ 1 \end{pmatrix} \Leftrightarrow UC^{II} = \{2, 3, 4, 5, 6\},$$

$$\mathbf{uc^{III}} = \overline{\overline{(\mathbf{T \cdot M + M \cdot M + M + T + E}) \cdot \mathbf{a}}} =$$

$$= \overline{\overline{\begin{pmatrix} 0 & 0 & 0 & 0 & 0 & 0 \\ 0 & 0 & 0 & 1 & 0 & 0 \\ 0 & 0 & 0 & 0 & 0 & 0 \\ 0 & 0 & 0 & 0 & 0 & 0 \\ 0 & 0 & 0 & 0 & 0 & 0 \\ 0 & 0 & 0 & 0 & 0 & 0 \end{pmatrix} \cdot \begin{pmatrix} 1 \\ 1 \\ 1 \\ 1 \\ 1 \\ 1 \end{pmatrix}}} = \overline{\begin{pmatrix} \overline{0} \\ \overline{1} \\ \overline{0} \\ \overline{0} \\ \overline{0} \\ \overline{0} \end{pmatrix}} = \begin{pmatrix} 1 \\ 0 \\ 1 \\ 1 \\ 1 \\ 1 \end{pmatrix} \Leftrightarrow UC^{III} = \{1, 3, 4, 5, 6\},$$

$$\mathbf{uc^{IV}} = \overline{\overline{(\mathbf{T \cdot M + M \cdot T + M \cdot M + M + T + E}) \cdot \mathbf{a}}} = \overline{\overline{\mathbf{I} \cdot \mathbf{a}}} = \overline{\overline{\mathbf{O} \cdot \mathbf{a}}} = \overline{\mathbf{o}} = \mathbf{a} \Leftrightarrow$$

$$\Leftrightarrow UC^{IV} = \{1, 2, 3, 4, 5, 6\} = A.$$

Evidently, $\mathbf{T^2 + I = I}$, therefore $\mathbf{uc^V} = \overline{\overline{(\mathbf{T \cdot T + T \cdot M + M \cdot T + M \cdot M + M + T + E}) \cdot \mathbf{a}}} = \overline{\overline{\mathbf{I} \cdot \mathbf{a}}} = \mathbf{a} \Leftrightarrow$

$UC^V = UC^{IV} = A$.

Finally, it should be noted that in terms of matrices **M** and **U** the expressions for the characteristic vectors can be written simpler: **uc**$^I$=$\overline{(\overline{M \cdot M + U}) \cdot a}$, **uc**$^{II}$=$\overline{(\overline{M \cdot U + U}) \cdot a}$, **uc**$^{III}$=$\overline{(\overline{U \cdot M + U}) \cdot a}$, **uc**$^{IV}$=$\overline{(\overline{U \cdot M + M \cdot U + U}) \cdot a}$ and **uc**$^V$=$\overline{\overline{U \cdot U} \cdot a}$.

*The uncaptured set*

The concept of the uncaptured set was proposed by Duggan (2007).

Let us define the capturing relation β. An alternative i is captured by an alternative j, (j, i)∈β iff none of the following propositions holds: 1) (j, i)∉µ; 2) ∃k: (iµk & kµj)∨(iµk & kτj)∨(iτk & kµj); 3) ∃k, l: (iµk & kµl & lµj)∨(iµk & kτl & lµj). An uncaptured alternative is an alternative, which is not captured by any alternative from A. The uncaptured set UCp is comprised of those and only those alternatives that are uncaptured, that is i∈UCp iff any alternative dominating i is either 1) reachable from i in two steps, at list one of which is a µ-step, or 2) reachable from i in three steps, the first and the last of which are µ-steps. Since capturing relation is asymmetric, uncaptured alternatives and only they are maximal elements of β, that is UCp=MAX(β).

Let **Q**=**M**·**T**·**M**+**M**·**M**·**M**+**M**·**T**+**T**·**M**+**M**·**M**+**M**+**T**+**E**. If $q_{ij}$=1 then either i=j, or iτj, or iµj, or ∃k: (iµk & kµj)∨(iµk & kτj)∨(iτk & kµj), or ∃k, l: (iµk & kµl & lµj)∨(iµk & kτl & lµj) hold. Consequently, if $q_{ij}$=1 then i is not captured by j, (j, i)∉β, and (j, i)∈β if $q_{ij}$=0. Therefore **R**=$\overline{Q}^{tr}$=$\overline{(M \cdot T \cdot M + M \cdot M \cdot M + T \cdot M + M \cdot T + M \cdot M + M + T + E)^{tr}}$ is a matrix representation of the capturing relation β.

Since capturing relation is asymmetric, by Lemma 2 we obtain the following formula for characteristic vector **ucp** of the uncaptured set UCp:

**ucp**=**max**(β)=$\overline{\overline{R}^{tr} \cdot a}$=$\overline{\overline{Q} \cdot a}$=$\overline{(M \cdot T \cdot M + M \cdot M \cdot M + T \cdot M + M \cdot T + M \cdot M + M + T + E) \cdot a}$,

**ucp**=$\overline{(M \cdot U \cdot M + U \cdot M + M \cdot U + U) \cdot a}$.

In our example **T**·**M**+**M**·**T**+**M**$^2$+**M**+**T**+**E**=**I**, therefore **ucp**=$\overline{\overline{I} \cdot a}$=**a** ⇔ UCp=UC$^{IV}$=A.

*The minimal dominant, minimal undominated and untrapped sets*

A set B, B⊆A, is called a dominant set if each alternative in B dominates each alternative outside B, $i \in B \Leftrightarrow (\forall j \in A \backslash B \Rightarrow i\mu j)$ (Ward, 1961; Smith, 1973). A dominant set will be called a minimal dominant set (denoted MD) if none of its proper subsets is a dominant set (Fishburn, 1977; Miller, 1977; Schwartz, 1977).

A set B, B⊆A, is called an undominated set if no alternative outside B dominates some alternative in B, $i \in B \Leftrightarrow (\forall j \in A \backslash B \Rightarrow (j, i) \notin \mu)$ (Ward, 1961). An undominated set is called a minimal undominated set if none of its proper subsets is an undominated set (Schwartz, 1970). If such a set is not unique, then the solution is defined as a union of these sets (Schwartz, 1972), which is denoted MU.

Let us define the trapping relation γ. It is said that an alternative i traps an alternative j iff i dominates j and i is not reachable from j via μ (Duggan, 2007), $(i, j) \in \gamma \Leftrightarrow (i, j) \in \mu \ \& \ (j, i) \notin \kappa(\mu)$. An untrapped set UT is comprised of those and only those alternatives that are not trapped by any alternative from A (Duggan, 2007). Since the trapping relation γ is asymmetric, untrapped alternatives and only they are maximal elements of γ, that is UT=MAX(γ).

Let us consider matrices $\mathbf{M}_{(k)} = \sum_{i=1}^{k} \mathbf{M}^i + \mathbf{E}$ and $\mathbf{U}_{(k)} = \sum_{i=1}^{k} \mathbf{U}^i$; $m_{(k)ij}=1$ iff there is a μ-path from an alternative i to an alternative j, j≠i, of length l: 0<l≤k, also $m_{(k)ii}=1$ for all i∈A. Consequently $\mathbf{M}_{(k)}$ is a matrix representing a k-transitive closure of μ, $\kappa_k(\mu)$. Respectively, $\mathbf{U}_{(k)}$ is a matrix representing $\kappa_k(\upsilon)$. Since A is finite there must be a minimal μ-path of length maximal among all minimal μ-paths in A. Let d(μ) denote the length of such a path, i.e. d(μ) is a μ-diameter of A. Then if follows from the definition of $\kappa_k(\rho)$ that $\kappa_k(\mu) \neq \kappa_{d(\mu)}(\mu)$ if k<d(μ) and $\kappa_{d(\mu)}(\mu) = \kappa_k(\mu) = \kappa(\mu)$ for any k: k≥d(μ). There must be a minimal υ-path of length maximal among all minimal υ-paths in A as well. Let its length be denoted by d(υ). Analogously, $\kappa_k(\upsilon) \neq \kappa_{d(\upsilon)}(\upsilon)$ if k<d(υ) and $\kappa_{d(\upsilon)}(\upsilon) = \kappa_k(\upsilon) = \kappa(\upsilon)$ for any k: k≥d(υ). Since $\mathbf{M}_{(k)}$ and $\mathbf{U}_{(k)}$ represent $\kappa_k(\mu)$ and $\kappa_k(\upsilon)$, $\mathbf{M}_{(d(\mu))}$ and $U_{(d(\upsilon))}$ are the representations of $\kappa(\mu)$ and $\kappa(\upsilon)$, respectively. Let us note that $\mathbf{M}_{(d(\mu))} \neq \mathbf{M}_{(d(\mu)-1)}$ & $\mathbf{M}_{(d(\mu))} = \mathbf{M}_{(d(\mu)+1)}$ and $\mathbf{U}_{(d(\upsilon))} \neq \mathbf{U}_{(d(\upsilon)-1)}$ & $\mathbf{U}_{(d(\upsilon))} = \mathbf{U}_{(d(\upsilon)+1)}$ hold.

To calculate the minimal dominant set MD and the union of minimal undominated sets MU we will use the following Theorem (Deb, 1977): $MU = MAX(\kappa(\mu))$, $MD = MAX(\kappa(\upsilon))$.

Since $MU = MAX(\kappa(\mu))$, $\rho = \kappa(\mu)$, $\mathbf{R} = \mathbf{M}_{(d)}$, $d = d(\mu)$. By Lemma 2 a characteristic vector of the union of minimal undominated sets MU is

$$\mathbf{mu} = \overline{\overline{(\mathbf{M}_{(d)} + \overline{\mathbf{M}_{(d)}^{tr}}) \cdot \mathbf{a}}}.$$

A diameter d is determined by a condition $\mathbf{M}_{(d)} \neq \mathbf{M}_{(d-1)}$ & $\mathbf{M}_{(d)} = \mathbf{M}_{(d+1)}$.

Now let us consider MD. Since $MD = MAX(\kappa(\upsilon))$, $\rho = \kappa(\mu)$, $\mathbf{R} = \mathbf{U}_{(d)}$, $d = d(\upsilon)$. Since $\upsilon$ is always complete and $\rho \subseteq \kappa_\kappa(\rho)$ holds for any natural k, completeness of $\upsilon$ implies completeness of $\kappa_\kappa(\upsilon)$ and $\kappa(\upsilon)$. Therefore by Lemma 2 a characteristic vector of the minimal dominant set MD is

$$\mathbf{md} = \overline{\overline{(\mathbf{U}_{(d)} + \mathbf{E}) \cdot \mathbf{a}}} = \overline{\overline{\mathbf{U}_{(d)} \cdot \mathbf{a}}}.$$

A diameter d is determined by a condition $\mathbf{U}_{(d)} \neq \mathbf{U}_{(d-1)}$ & $\mathbf{U}_{(d)} = \mathbf{U}_{(d+1)}$.

Let $\mathbf{Q} = \mathbf{M}_{(d(\mu))} + \mathbf{T}$. If $q_{ij} = 1$ then either $i\tau j$ or $i\kappa(\mu)j$ holds. Consequently, if $q_{ij} = 1$ then i is not trapped by j, $(j, i) \notin \gamma$, and $(j, i) \in \gamma$ if $q_{ij} = 0$. Therefore if $d = d(\mu)$ then $\mathbf{R} = \overline{\mathbf{Q}^{tr}} = \overline{(\mathbf{M}_{(d)} + \mathbf{T})^{tr}}$ is a matrix representation of the trapping relation $\gamma$.

Since trapping is asymmetric, by Lemma 2 we obtain the following formula for the characteristic vector $\mathbf{ut}$ of the untrapped set UT:

$$\mathbf{ut} = \mathbf{max}(\gamma) = \overline{\overline{\mathbf{R}^{tr} \cdot \mathbf{a}}} = \overline{\overline{\mathbf{Q} \cdot \mathbf{a}}} = \overline{\overline{(\mathbf{M}_{(d)} + \mathbf{T}) \cdot \mathbf{a}}}.$$

A diameter $d = d(\mu)$ is determined by a condition $\mathbf{M}_{(d)} \neq \mathbf{M}_{(d-1)}$ & $\mathbf{M}_{(d)} = \mathbf{M}_{(d+1)}$ or, alternatively, $\mathbf{M}^d \neq \mathbf{M}_{(d-1)}$ & $\mathbf{M}^{d+1} = \mathbf{M}_{(d)}$

For the example considered above we obtain

$$\mathbf{M}_{(2)}=\mathbf{M}^2+\mathbf{M}_{(1)}=\begin{pmatrix}0&0&1&0&0&0\\1&0&0&0&0&0\\0&1&0&0&0&0\\0&1&1&0&0&1\\0&1&0&1&0&0\\1&1&1&0&1&0\end{pmatrix}+\begin{pmatrix}1&1&0&0&0&0\\0&1&1&0&0&0\\1&0&1&0&0&0\\1&1&0&1&1&0\\0&0&0&0&1&1\\0&1&0&1&0&1\end{pmatrix}=\begin{pmatrix}1&1&1&0&0&0\\1&1&1&0&0&0\\1&1&1&0&0&0\\1&1&1&1&1&1\\0&1&0&1&1&1\\1&1&1&1&1&1\end{pmatrix},$$

$$\mathbf{M}_{(3)}=\mathbf{M}^3+\mathbf{M}_{(2)}=\begin{pmatrix}1&0&0&0&0&0\\0&1&0&0&0&0\\0&0&1&0&0&0\\1&1&1&1&0&0\\1&1&1&0&1&0\\1&1&1&0&0&1\end{pmatrix}+\begin{pmatrix}1&1&1&0&0&0\\1&1&1&0&0&0\\1&1&1&0&0&0\\1&1&1&1&1&1\\0&1&0&1&1&1\\1&1&1&1&1&1\end{pmatrix}=\begin{pmatrix}1&1&1&0&0&0\\1&1&1&0&0&0\\1&1&1&0&0&0\\1&1&1&1&1&1\\1&1&1&1&1&1\\1&1&1&1&1&1\end{pmatrix}\ne\mathbf{M}_{(2)},$$

$$\mathbf{M}_{(4)}=\mathbf{M}^4+\mathbf{M}_{(3)}=\begin{pmatrix}1&1&1&0&0&0\\1&1&1&0&0&0\\1&1&1&0&0&0\\1&1&1&1&1&1\\1&1&1&1&1&1\\1&1&1&1&1&1\end{pmatrix}+\begin{pmatrix}1&1&1&0&0&0\\1&1&1&0&0&0\\1&1&1&0&0&0\\1&1&1&1&1&1\\1&1&1&1&1&1\\1&1&1&1&1&1\end{pmatrix}=\begin{pmatrix}1&1&1&0&0&0\\1&1&1&0&0&0\\1&1&1&0&0&0\\1&1&1&1&1&1\\1&1&1&1&1&1\\1&1&1&1&1&1\end{pmatrix}=\mathbf{M}_{(3)}.$$

Therefore d(μ)=3.

$$\mathbf{mu}=\overline{\overline{(\mathbf{M}_{(3)}+\overline{\mathbf{M}}_{(3)}^{\text{tr}})}\cdot\mathbf{a}}=$$

$$=\overline{\overline{\left(\begin{pmatrix}1&1&1&0&0&0\\1&1&1&0&0&0\\1&1&1&0&0&0\\1&1&1&1&1&1\\1&1&1&1&1&1\\1&1&1&1&1&1\end{pmatrix}+\begin{pmatrix}0&0&0&0&0&0\\0&0&0&0&0&0\\0&0&0&0&0&0\\1&1&1&0&0&0\\1&1&1&0&0&0\\1&1&1&0&0&0\end{pmatrix}\right)\cdot\begin{pmatrix}1\\1\\1\\1\\1\\1\end{pmatrix}}}=\overline{\begin{pmatrix}0&0&0&1&1&1\\0&0&0&1&1&1\\0&0&0&1&1&1\\0&0&0&0&0&0\\0&0&0&0&0&0\\0&0&0&0&0&0\end{pmatrix}\cdot\begin{pmatrix}1\\1\\1\\1\\1\\1\end{pmatrix}}=\begin{pmatrix}0\\0\\0\\1\\1\\1\end{pmatrix}.$$

Consequently, MU={4, 5, 6}.

$$\mathbf{ut}=\overline{\overline{(\mathbf{M}_{(3)}+\mathbf{T})\cdot\mathbf{a}}}=$$

$$
=\overline{\left(\begin{pmatrix} 1 & 1 & 1 & 0 & 0 & 0 \\ 1 & 1 & 1 & 0 & 0 & 0 \\ 1 & 1 & 1 & 0 & 0 & 0 \\ 1 & 1 & 1 & 1 & 1 & 1 \\ 1 & 1 & 1 & 1 & 1 & 1 \\ 1 & 1 & 1 & 1 & 1 & 1 \end{pmatrix} + \begin{pmatrix} 0 & 0 & 0 & 0 & 1 & 1 \\ 0 & 0 & 0 & 0 & 1 & 0 \\ 0 & 0 & 0 & 1 & 1 & 1 \\ 0 & 0 & 1 & 0 & 0 & 0 \\ 1 & 1 & 1 & 0 & 0 & 0 \\ 1 & 0 & 1 & 0 & 0 & 0 \end{pmatrix}\right)} \cdot \begin{pmatrix} 1 \\ 1 \\ 1 \\ 1 \\ 1 \\ 1 \end{pmatrix} = \overline{\begin{pmatrix} 0 & 0 & 0 & 1 & 0 & 0 \\ 0 & 0 & 0 & 1 & 0 & 1 \\ 0 & 0 & 0 & 0 & 0 & 0 \\ 0 & 0 & 0 & 0 & 0 & 0 \\ 0 & 0 & 0 & 0 & 0 & 0 \\ 0 & 0 & 0 & 0 & 0 & 0 \end{pmatrix}} \cdot \begin{pmatrix} 1 \\ 1 \\ 1 \\ 1 \\ 1 \\ 1 \end{pmatrix} = \begin{pmatrix} 0 \\ 0 \\ 1 \\ 1 \\ 1 \\ 1 \end{pmatrix}
$$

Consequently, UT={3, 4, 5, 6}=UC$^I$.

Note that MD=A since always UCp⊆MD (Duggan, 2007) and UCp=A in this example.

### *The minimal weakly stable set*

The first version of this solution was introduced by Aleskerov and Kurbanov (1999). A set B, B⊆A, is called a weakly stable set if it has the following property: if i belongs to B, then for any j outside B, which dominates i, there is an alternative k in B, which dominates j, $\forall i \in A$, $i \in B \Leftrightarrow (\exists j \notin B: j\mu i \Rightarrow \exists k \in B: k\mu j)$. In terms of upper and lower contour sets B is weakly stable iff $\forall j \notin B$ B∩L(j)≠∅ ⇒ B∩D(j)≠∅.

Subochev (2008) proposed a second version of a weakly stable set. B is a weakly stable set iff $\forall j$: $j \notin B \Rightarrow$ B∩D(j)≠∅. That is B is a weakly stable set iff there is one-step path from some alternative in B to any alternative outside B.

A weakly stable set is called a minimal weakly stable set if none of its proper subsets is a weakly stable set. If such set is not unique, then the solution is defined as a union of these sets. Thus we have two versions of this solution: MWS$^I$ and MWS$^{II}$.

To calculate a second version of a union of minimal weakly stable sets MWS$^{II}$ we will use the following Theorem (see Subochev (2008)): an alternative i belongs to a minimal weakly stable set (second version) iff i is uncovered according to the third version of the covering relation or some alternative from the lower contour set of i is uncovered according to the third version of the covering relation, $i \in MWS^{II} \Leftrightarrow$ either $i \in UC^{III}$, or $\exists j$: $j \in L(i)$ & $j \in UC^{III}$. That is an alternative belongs to a union of minimal weakly stable sets MWS$^{II}$ iff it ether belongs to UC$^{III}$, or belongs to an upper contour of some alternative from UC$^{III}$. Consequently, MWS$^{II}$ is a union of UC$^{III}$ and upper contours of all alternatives from $UC^{III}$. A characteristic vector $\mathbf{d}(i)$ of upper contour set of an alternative i is given by the formula $\mathbf{d}(i)=\mathbf{M}\cdot\mathbf{e}(i)$. Therefore, for a characteristic vector $\mathbf{d}(UC^{III})$ of a union of upper contour sets of all alternatives from $UC^{III}$ we obtain

$$\mathbf{d}(UC^{III})= \sum_{i\in UC^{III}}\mathbf{d}(i) = \sum_{i\in UC^{III}}\mathbf{M}\cdot\mathbf{e}(i) =\mathbf{M}\cdot \sum_{i\in UC^{III}}\mathbf{e}(i) =\mathbf{M}\cdot\mathbf{uc}^{III}.$$

Thus $\mathbf{mws}^{II}=\mathbf{uc}^{III}+\mathbf{d}(UC^{III})=\mathbf{uc}^{III}+\mathbf{M}\cdot\mathbf{uc}^{III}=(\mathbf{M}+\mathbf{E})\cdot\mathbf{uc}^{III}$ and finally

$$\mathbf{mws}^{II}=(\mathbf{M}+\mathbf{E})\cdot\overline{(\mathbf{T}\cdot\mathbf{M}+\mathbf{M}\cdot\mathbf{M}+\mathbf{M}+\mathbf{T}+\mathbf{E})\cdot\mathbf{a}}=(\mathbf{M}+\mathbf{E})\cdot\overline{(\mathbf{U}\cdot\mathbf{M}+\mathbf{U})\cdot\mathbf{a}}.$$

In our example $\mathbf{uc}^{III}=\begin{pmatrix}1\\0\\1\\1\\1\\1\end{pmatrix} \Rightarrow \mathbf{mws}^{II}=\begin{pmatrix}1&1&0&0&0&0\\0&1&1&0&0&0\\1&0&1&0&0&0\\1&1&0&1&1&0\\0&0&0&0&1&1\\0&1&0&1&0&1\end{pmatrix}\cdot\begin{pmatrix}1\\0\\1\\1\\1\\1\end{pmatrix}=\begin{pmatrix}1\\1\\1\\1\\1\\1\end{pmatrix}=\mathbf{a} \Leftrightarrow MWS^{III}=A.$

Unfortunately, we cannot get similar representation for the original version of a union of minimal weakly stable sets $MWS^{I}$.

**4. New versions of the uncovered and weakly stable sets**

As it has been already noted in Subochev (2008) neither Fishburn, nor Miller did explicitly include a condition $j\mu i$ into their definitions of the covering relation (third and second versions, respectively). That is none of them says that an alternative i can not be covered by an alternative j, which ties i, $j\tau i$. Miller (1980, p. 94) proposed only $L(i)\subseteq L(j)$ & $D(j)\subseteq D(i)$ as a definition of the covering relation for general case, $\tau\neq\varnothing$. If $\mu$ is a tournament then it does not matter which version of the covering relation is applied, since in tournaments Miller's and Fishburn's versions coincide with all other versions and imply $j\mu i$ when j covers i. But if there are ties, $\tau\neq\varnothing$, the absence of this condition in the definition makes a difference. If it is dropped one gets five more versions of the covering relation and of the uncovered set.

1) j covers i if $L(i)\subseteq L(j)\cup H(j)$, i is uncovered $\Leftrightarrow \forall j: j\neq i \Rightarrow i\mu j$ or $\exists k: i\mu k$ & $k\mu j$;

2) j covers i if $L(i)\subseteq L(j)$ (Miller, 1980),

i is uncovered $\Leftrightarrow \forall j: j \neq i \Rightarrow i\mu j$ or $\exists k: (i\mu k \ \& \ k\mu j) \vee (i\mu k \ \& \ k\tau j)$;

3) j covers i if $D(j) \subseteq D(i)$ (Fishburn, 1977),

i is uncovered $\Leftrightarrow \forall j: j \neq i \Rightarrow i\mu j$ or $\exists k: (i\mu k \ \& \ k\mu j) \vee (i\tau k \ \& \ k\mu j)$;

4) j covers i if $L(i) \subseteq L(j) \ \& \ D(j) \subseteq D(i)$ (Miller, 1980),

i is uncovered $\Leftrightarrow \forall j: j \neq i \Rightarrow i\mu j$ or $\exists k: (i\mu k \ \& \ k\mu j) \vee (i\mu k \ \& \ k\tau j) \vee (i\tau k \ \& \ k\mu j)$;

5) j covers i if $H(i) \cup L(i) \subseteq L(j)$,

i is uncovered $\Leftrightarrow \forall j: j \neq i \Rightarrow i\mu j$ or $\exists k: (i\mu k \ \& \ k\mu j) \vee (i\mu k \ \& \ k\tau j) \vee (i\tau k \ \& \ k\mu j) \vee (i\tau k \ \& \ k\tau j)$.

The condition $j\mu i$ in the definitions is what makes the covering relation asymmetric. Under these modified definitions the covering relation may possess a symmetric component. For instance, if $i\tau j$ and $L(i)=L(j)$ then i covers j and j covers i according to Miller's definition of the covering relation.

Let $\alpha^N_m$, $UC^N_m$, $\mathbf{uc}^N_m$ (N=I÷V) denote the modified versions of the covering relation, the corresponding versions of the uncovered set and their characteristic vectors, respectively. It follows from the definitions that modified uncovered sets are smaller than original ones: $UC^N_m \subseteq UC^N$. Considerations similar to those that produced matrix-vector representation of $\alpha^I_m$ and $UC^I$ lead us to the following formulae

$\rho = \alpha^I_m \Rightarrow \mathbf{R} = \overline{(\mathbf{M} \cdot \mathbf{M} + \mathbf{M} + \mathbf{E})^{tr}}$, $\mathbf{uc}^I_m = \overline{(\mathbf{M} \cdot \mathbf{M} + \mathbf{M} + \mathbf{E}) \cdot \mathbf{a}}$;

$\rho = \alpha^{II}_m \Rightarrow \mathbf{R} = \overline{(\mathbf{M} \cdot \mathbf{T} + \mathbf{M} \cdot \mathbf{M} + \mathbf{M} + \mathbf{E})^{tr}}$, $\mathbf{uc}^{II}_m = \overline{(\mathbf{M} \cdot \mathbf{T} + \mathbf{M} \cdot \mathbf{M} + \mathbf{M} + \mathbf{E}) \cdot \mathbf{a}}$;

$\rho = \alpha^{III}_m \Rightarrow \mathbf{R} = \overline{(\mathbf{T} \cdot \mathbf{M} + \mathbf{M} \cdot \mathbf{M} + \mathbf{M} + \mathbf{E})^{tr}}$, $\mathbf{uc}^{III}_m = \overline{(\mathbf{T} \cdot \mathbf{M} + \mathbf{M} \cdot \mathbf{M} + \mathbf{M} + \mathbf{E}) \cdot \mathbf{a}}$;

$\rho = \alpha^{IV}_m \Rightarrow \mathbf{R} = \overline{(\mathbf{T} \cdot \mathbf{M} + \mathbf{M} \cdot \mathbf{T} + \mathbf{M} \cdot \mathbf{M} + \mathbf{M} + \mathbf{E})^{tr}}$, $\mathbf{uc}^{IV}_m = \overline{(\mathbf{T} \cdot \mathbf{M} + \mathbf{M} \cdot \mathbf{T} + \mathbf{M} \cdot \mathbf{M} + \mathbf{M} + \mathbf{E}) \cdot \mathbf{a}}$;

$\rho = \alpha^V_m \Rightarrow \mathbf{R} = \overline{(\mathbf{T} \cdot \mathbf{T} + \mathbf{T} \cdot \mathbf{M} + \mathbf{M} \cdot \mathbf{T} + \mathbf{M} \cdot \mathbf{M} + \mathbf{M} + \mathbf{E})^{tr}}$,

$\mathbf{uc}^V_m = \overline{(\mathbf{T} \cdot \mathbf{T} + \mathbf{T} \cdot \mathbf{M} + \mathbf{M} \cdot \mathbf{T} + \mathbf{M} \cdot \mathbf{M} + \mathbf{M} + \mathbf{E}) \cdot \mathbf{a}}$.

Absence of $\mathbf{T}$ in a sum is what differs all of these formulae from their unmodified counterparts. In terms of $\mathbf{M}$ and $\mathbf{U}$ the expressions for the characteristic vectors are the following:

$$\mathbf{uc}^{II}{}_m = \overline{\overline{(\mathbf{M}\cdot\mathbf{U}+\mathbf{E})\cdot\mathbf{a}}},$$

$$\mathbf{uc}^{III}{}_m = \overline{\overline{(\mathbf{U}\cdot\mathbf{M}+\mathbf{E})\cdot\mathbf{a}}},$$

$$\mathbf{uc}^{IV}{}_m = \overline{\overline{(\mathbf{U}\cdot\mathbf{M}+\mathbf{M}\cdot\mathbf{U}+\mathbf{E})\cdot\mathbf{a}}},$$

$$\mathbf{uc}^{V}{}_m = \overline{\overline{(\mathbf{T}\cdot\mathbf{T}+\mathbf{U}\cdot\mathbf{M}+\mathbf{M}\cdot\mathbf{U}+\mathbf{E})\cdot\mathbf{a}}}.$$

Let us also propose a new (third) version for the definition of a weakly stable set: B, B⊆A, is a weakly stable set iff ∀i: i∈A\B ⇒ B∩(D(i)∪H(i))≠∅. That is B is a weakly stable set iff there is a one-step υ-path from some alternative in B to any alternative outside B, ∀i: i∈A\B ∃ j: j∈B & jυi. Correspondingly, B is not a weakly stable set iff ∃i: B⊆L(i), i.e. iff there is an alternative dominating all alternatives from B.

The weak stability under the third definition, like its second version, is monotonous. That is if B⊆C and B is a weakly stable set, then C is a weakly stable set as well. If C is not a weakly stable set, then any B, such that B⊆C, is not a weakly stable set.

This new definition gives us one more solution - a third version of a union of minimal weakly stable sets $MWS^{III}$. Minimality is defined here in a standard way: a set is a minimal weakly stable set if none of its proper subsets is weakly stable. The difference between sets $MWS^I$, $MWS^{II}$ and $MWS^{III}$ lies in the definition of the weak stability only. In a tournament $MWS^{III}$ coincides with $MWS^I$ and $MWS^{II}$.

A criterion to determine whether an alternative belongs to a minimal weakly stable set under the third definition is given by the following

**Lemma 3.** An alternative i belongs to a minimal weakly stable set (third version) $MWS^{III}$ iff i is uncovered according to the second version of the covering relation or some alternative from the lower contour set of i or from the horizon of i is not covered by any alternative from the upper contour set of i according to the modified second version of the covering relation, i∈$MWS^{III}$ ⇔ either i∈$UC^{II}$, or ∃j: j∈L(i)∪H(i) & (∀k: k∈D(i) ⇒ (jμk or ∃l: (jμl & lμk)∨(jμl & lτk))). Correspondingly i∉$MWS^{III}$ ⇔ (i∉$UC^{II}$ and ∀j: j∈L(i)∪H(i) ⇒ ∃k: k∈D(i) & L(j)⊆L(k)).

The proof of Lemma 3 is given in the Appendix.

Lemma 3 allows us to find a matrix representation for $MWS^{III}$. Let $\mathbf{R} = \overline{\mathbf{M} \cdot \mathbf{T} + \mathbf{M} \cdot \mathbf{M} + \mathbf{M} + \mathbf{E}} = \overline{\mathbf{M} \cdot \mathbf{U} + \mathbf{E}}$. Then $r_{ij}=0$ iff i is not covered by j according to the modified second version of the covering relation. Let **b** and **c** be characteristic vectors of sets B and C, $B \subseteq A$, $C \subseteq A$, respectively. Let $\mathbf{v} = \mathbf{R} \cdot \mathbf{b}$, then $v_i=1$ iff an alternative i is covered at list by one alternative from the set B according to the modified second version of the covering relation. Consequently, $\overline{v}_i = 1$ iff an alternative i is not covered by any alternative from the set B according to the modified second version of the covering relation. Then a scalar product $(\mathbf{c} \cdot \overline{\mathbf{v}}) = \sum_{k=1}^{n} c_k \cdot \overline{v}_k = 1$ iff there is at list one alternative in C not covered by any alternative from the set B according to the modified second version of the covering relation. Now let $B=D(i)$, $C=L(i) \cup H(i)$ and $f_i = (\mathbf{c} \cdot \overline{\mathbf{v}})$. Then $f_i=1$ iff there is some alternative from the lower contour set of i or from the horizon of i not covered by any alternative from the upper contour of i according to the modified second version of the covering relation, i.e. **f** is a characteristic vector of precisely those alternatives that satisfy the aforementioned condition. According to the formulae (1)

$b_j = d(i)_j = (\mathbf{M} \cdot \mathbf{e}(i))_j = m_{ji}$; $c_k = l(i)_k + h(i)_k = (\mathbf{M}^{tr} \cdot \mathbf{e}(i))_k + (\mathbf{T} \cdot \mathbf{e}(i))_k = m_{ik} + t_{ki} = m_{ik} + t_{ik}$.

As a result $f_i = \sum_{j,k=1}^{n} (m_{ik} + t_{ik}) \cdot \overline{(\mathbf{M} \cdot \mathbf{U} + \mathbf{E})}_{kj} \cdot m_{ji}$, that is $\mathbf{f} = \text{diag}((\mathbf{M}+\mathbf{T}) \cdot \overline{(\mathbf{M} \cdot \mathbf{U} + \mathbf{E})} \cdot \mathbf{M})$.

Let $\mathbf{mws}^{III}$ denote a characteristic vector of $MWS^{III}$. Then by Theorem 1 $\mathbf{mws}^{III} = \mathbf{uc}^{II} + \mathbf{f}$. Therefore $\mathbf{mws}^{III} = \overline{(\mathbf{M} \cdot \mathbf{U} + \mathbf{U})} \cdot \mathbf{a} + \text{diag}((\mathbf{M}+\mathbf{T}) \cdot \overline{(\mathbf{M} \cdot \mathbf{U} + \mathbf{E})} \cdot \mathbf{M})$.

**5. Classes of k-stable alternatives and k-stable sets**

Let μ be a tournament. Then $\tau = \varnothing$, $\upsilon = \mu$, $\mathbf{M}_{(k)} = \mathbf{U}_{(k)}$, $d(\mu) = d(\upsilon) = d$ and all steps and paths are μ-steps and μ-paths. Since μ is complete, all $\kappa_k(\mu)$ are complete as well. Some solutions considered above coincide in a tournament:

1) the Condorcet winner coincides with the core $\{CW\} = Cr$;

2) all versions of the uncovered sets coincide with each other and are denoted as UC;

3) all versions of the union of minimal weakly stable sets coincide with each other and are denoted as MWS;

4) the union of minimal undominated sets MU and the untrapped set UT coincide with the minimal dominant set MD, MU=UT=MD.

An alternative i is called generally stable if every other alternative in A is reachable from i. Every alternative in A is reachable from i iff i belongs to a minimal dominant set (Miller, 1977), thus all alternatives of a minimal dominant set and only they are generally stable. Since A is finite, if j is reachable from i, then there is a path from i to j with a minimal length. Let l(i, j) denote a minimal length function, i.e. l(i, j) is equal to the length of a minimal path from i to j. By definition l(i, i)=0.

An alternative i is called a k-stable alternative if $\max_{j \in A} l(i, j) = k$, i.e. if it is possible to reach any other alternative in A from i in no more than k steps, but there is at list one alternative reachable from i in less than k steps (Aleskerov, Subochev, 2009). $SP_{(k)}$ denotes a class of k-stable alternatives in A. $P_{(k)}$ denotes a set of those generally stable alternatives, from which it is possible to reach any given alternative in A in no more than k steps, $P_{(k)}=SP_{(1)}+SP_{(2)}+\ldots+SP_{(k)}$.

This definition and the fact that $\tau=\varnothing$ together imply $i \in P_{(k)} \Leftrightarrow i \in MAX(\kappa_k(\mu))$, i.e. $P_{(k)}=MAX(\kappa_k(\mu))$. Let $\mathbf{p}_{(k)}$ denote a characteristic vector of $P_{(k)}$. By Lemma 2 $\mathbf{p}_{(k)}= \overline{\overline{(\mathbf{M}_{(k)} + \mathbf{E}) \cdot \mathbf{a}}} = \overline{\overline{\mathbf{M}_{(k)} \cdot \mathbf{a}}}$. Since all matrices are Boolean ones, the following equation holds $\sum_{i=1}^{k} \mathbf{M}^i + \mathbf{E} = (\mathbf{M}+\mathbf{E})^k$. Consequently $\mathbf{M}_{(k)} = \sum_{i=1}^{k} \mathbf{M}^i + \mathbf{E} = \mathbf{U}^k$ and $\mathbf{p}_{(k)} = \overline{\overline{\mathbf{U}^k \cdot \mathbf{a}}}$.

Let $\mathbf{sp}_{(k)}$ denote a characteristic vector of $SP_{(k)}$. By definition of $P_{(k)}$ $i \in SP_{(k)} \Leftrightarrow i \in P_{(k)}$ & $i \notin P_{(k-1)}$. Therefore $sp_{(k)i}=p_{(k)i} \cdot \overline{p_{(k-1)i}} = \overline{\overline{p_{(k)i}} + p_{(k-1)i}}$, that is

$$\mathbf{sp}_{(k)} = \overline{\overline{\mathbf{p}_{(k)}} + \mathbf{p}_{(k-1)}} = \overline{\overline{\mathbf{M}_{(k)} \cdot \mathbf{a}} + \overline{\overline{\mathbf{M}_{(k-1)}}} \cdot \mathbf{a}} = \overline{\overline{\mathbf{U}^k \cdot \mathbf{a}} + \overline{\overline{\mathbf{U}^{k-1} \cdot \mathbf{a}}}}.$$

If k=1 then $\mathbf{M}_{(k-1)}=\mathbf{E}$. $\overline{\overline{\mathbf{E} \cdot \mathbf{a}}} = \mathbf{o} \Rightarrow \mathbf{sp}_{(1)}=\mathbf{p}_{(1)}= \overline{\overline{(\mathbf{M}+\mathbf{E}) \cdot \mathbf{a}}} = \mathbf{cw}$ - a Condorcet winner.

If k=2 then $\mathbf{M}_{(k-1)}=\mathbf{M}+\mathbf{E}$. $\overline{(\mathbf{M}+\mathbf{E})\cdot\mathbf{a}}=\mathbf{cw}$. If $\mathbf{cw}\neq\mathbf{o}$, i.e. if there is a Condorcet winner, then $\overline{\overline{(\mathbf{M}^2+\mathbf{M}+\mathbf{E})\cdot\mathbf{a}}}=\overline{\mathbf{cw}}$. $\mathbf{cw}+\overline{\mathbf{cw}}=\mathbf{a}$ $\Rightarrow$ $\mathbf{sp}_{(2)}=\overline{\overline{\mathbf{cw}}+\mathbf{cw}}=\overline{\mathbf{a}}=\mathbf{o}$, that is $SP_{(2)}$ is empty. If there is no Condorcet winner, $\mathbf{cw}=\mathbf{o}$, then $\mathbf{sp}_{(2)}=\mathbf{p}_{(2)}=\overline{\overline{(\mathbf{M}^2+\mathbf{M}+\mathbf{E})\cdot\mathbf{a}}}=\mathbf{uc}$, which corresponds to $SP_{(2)}=UC$ if $\{CW\}=\varnothing$.

It is also evident that $\mathbf{p}_{(3)}=\overline{\overline{(\mathbf{M}^3+\mathbf{M}^2+\mathbf{M}+\mathbf{E})\cdot\mathbf{a}}}=\mathbf{ucp}$.

Since A is finite there must be a finite number $m=\max_{i\in MD}(\max_{j\in A} l(i, j))$ - a maximum of degrees of stability, i.e. a maximum of lengths of minimal paths $l(i, j)$ from alternatives that belong to MD to all alternatives in A. Then $P_{(m)}=MD$ and $SP_{(k)}=\varnothing$ for all k: k>m.

By definition a µ-diameter of A, which was used in calculations of **mu** and **ut**, is a maximum of lengths of minimal paths $l(i, j)$ from *all* alternatives in A to all other alternatives in A: $d=\max_{i\in A}(\max_{j\in A} l(i, j))$. Consequently, $m=\max_{i\in MD}(\max_{j\in A} l(i, j))\leq\max_{i\in A}(\max_{j\in A} l(i, j))=d$. Since in a tournament $MU=UT=MD=P_{(m)}$, one needs not multiply matrices **U** till the value of d is determined - it is enough to find m (which might be much smaller than d) and then stop.

As it was shown in Subochev (2008) $SP_{(k)}\neq\varnothing$ for all k: k≤m and $SP_{(k)}=\varnothing$ for all k: k>m, the value of m can be determined from the condition $\mathbf{p}_{(m-1)}\neq\mathbf{p}_{(m)}$ & $\mathbf{p}_{(m)}=\mathbf{p}_{(m+1)}$, i.e.

$P_{(m)}=MD \Leftrightarrow \mathbf{md}=\mathbf{p}_{(m)}$.

A set B, B⊆A, is called a k-stable set if for any alternative j outside B, j∈A\B, there exists a µ-path of length l: l≤k to j from some alternative i from B, i∈B, but at the same time there is at list one alternative j outside B, j∈A\B, such that it is not reachable in less than k µ-steps from any i: i∈B (Subochev, 2009). A k-stable set will be called a minimal k-stable set if none of its proper subsets is a k-stable set. It follows from this definition that a weakly stable set is a 1-stable set.

$SS_{(k)}$ denotes a class of those alternatives, which belong to some minimal k-stable set, but do not belong to any minimal stable set with the degree of stability less than k. By construction these classes do not intersect. $S_{(k)}$ denotes a union of those minimal generally stable sets, from which it is possible to reach any alternative outside a set in no more than k steps. Evidently $S_{(k)}=SS_{(1)}+SS_{(2)}+\ldots+SS_{(k)}$.

A relation $\mu$ is asymmetric, but if there is no Condorcet winner, all relations $\kappa_k(\mu)$, $k\geq 2$, possess a symmetric component, since all $\kappa_k(\mu)$ are complete and $|MAX(\kappa_2(\mu))|=|P_{(2)}|=|UC|\geq 3$ for any A: $|A|\geq 4$ (Miller, 1980). Let $\upsilon_{(k)}$ denote $\kappa_k(\mu)$, $\upsilon_{(k)}= \kappa_k(\mu)$. Let $\mu_{(k)}=\pi(\kappa_k(\mu))$ and $\tau_{(k)}=\sigma(\kappa_k(\mu))$. It follows from the definitions that $\upsilon_{(1)}=\upsilon$, $\mu_{(1)}=\mu$, $\tau_{(1)}=\tau\cup\varepsilon$.

Let us consider $\upsilon_{(k)}$ and $\mu_{(k)}$ as new versions of relations $\upsilon$ and $\mu$. If a set B, $B\subseteq A$, is a k-stable set it follows from the definition of a k-stable set that any alternative j outside B will be reachable from some alternative i from B in one $\upsilon_{(k)}$-step, i.e. $\forall j: j\in A\setminus B \Rightarrow \exists i: i\in B$ & $i\upsilon_{(k)}j$. If the degree of stability of B is greater than k, then $\exists j: j\in A\setminus B$ & $\forall i: i\in B \Rightarrow (i, j)\notin \upsilon_{(k)}$. Consequently, if B is a minimal k-stable set with respect to $\mu$, it must be a minimal weakly stable set (third version) with respect to $\upsilon_{(k)}$. Conversely, if B is a minimal weakly stable set (third version) with respect to $\upsilon_{(k)}$, then it must be a minimal stable set with degree of stability no less than k with respect to $\mu$.

Let $\mathbf{ss}_{(k)}$ and $\mathbf{s}_{(k)}$ denote characteristic vectors for classes of k-stable sets $SS_{(k)}$ and their sums $S_{(k)}=SS_{(1)}+SS_{(2)}+\ldots+SS_{(k)}$, respectively. Let $MWS^{III}(\upsilon_{(k)})$ and $\mathbf{mws}^{III}(\upsilon_{(k)})$ denote a union of minimal weakly stable sets (third version) and its characteristic vector calculated with respect to the relation $\upsilon_{(k)}$ on A. Then $i\in SS_{(k)} \Rightarrow i\in MWS^{III}(\upsilon_{(k)})$ and $i\in MWS^{III}(\upsilon_{(k)}) \Rightarrow i\in SS_{(x)}$, x: $x\leq k$, that is $SS_{(k)}\subseteq MWS^{III}(\upsilon_{(k)})$ and $MWS^{III}(\upsilon_{(k)})\subseteq S_{(k)}$. Consequently $\mathbf{s}_{(k)}=\mathbf{mws}^{III}(\upsilon_{(k)})+\mathbf{s}_{(k-1)}$. Since the first class $SS_{(1)}$ is nothing other than a union of weakly stable sets (with respect to $\mu$), we obtain the following inductive formulae for calculation of $\mathbf{s}_{(k)}$:

$$\mathbf{s}_{(1)}=\mathbf{ss}_{(1)}=\mathbf{mws}=(\mathbf{M}+\mathbf{E})\cdot\mathbf{p}_{(2)}=\mathbf{U}\cdot\overline{\overline{\mathbf{U}^2}\cdot\mathbf{a}},$$

$$\mathbf{s}_{(k)}=\mathbf{s}_{(k-1)}+\mathbf{mws}^{III}(\upsilon_{(k)})=\mathbf{s}_{(k-1)}+\overline{\overline{(\tilde{\mathbf{M}}\cdot\tilde{\mathbf{U}}+\tilde{\mathbf{U}})\cdot\mathbf{a}}} +\mathrm{diag}(\,(\tilde{\mathbf{M}}+\tilde{\mathbf{T}})\cdot\overline{\overline{(\tilde{\mathbf{M}}\cdot\tilde{\mathbf{U}}+\mathbf{E})}}\cdot\tilde{\mathbf{M}}\,),$$

where $\tilde{\mathbf{U}}=\mathbf{M}_{(k)}$, $\tilde{\mathbf{M}}=\overline{(\mathbf{M}_{(k)}^{\mathrm{tr}}+\overline{\mathbf{M}}_{(k)})}$.

Since $P_{(k)} \subseteq S_{(k)} \subseteq P_{(k+2)} \subseteq MD$ (Subochev, 2009) iterrations will stop somewhere between k=m-2 and k=m, when $\mathbf{s}_{(k)}$ becomes equal to $\mathbf{md}=\mathbf{p}_{(m)}$. Finally $i \in SS_{(k)} \Leftrightarrow i \in S_{(k)}$ & $i \notin S_{(k-1)}$. Therefore $ss_{(k)i} = s_{(k)i} \cdot \overline{s_{(k-1)i}} = \overline{\overline{s_{(k)i}} + s_{(k-1)i}}$, that is $\mathbf{ss}_{(k)} = \overline{\overline{\mathbf{s}_{(k)}} + \mathbf{s}_{(k-1)}}$.

## 6. Conclusion

The following Theorem summarizes the results of this paper.

**Theorem.** Let **cw**, **cr**, $\mathbf{uc}^N$ (N=I÷V) and $\mathbf{uc}^N_m$, $\mathbf{mws}^{II}$, $\mathbf{mws}^{III}$, **ucp**, **mu**, **ut** and **md**, respectively, denote characteristic vectors of the following solutions: the Condorcet winner {CW}, the core Cr, five versions of the uncovered set $UC^N$ (N=I÷V) and their modifications $UC^N_m$, the second and the third versions of the union of minimal weakly stable sets $MWS^{II}$ and $MWS^{III}$, the uncaptured set UCp, the union of minimal undominated sets (strong top-cycles) MU, the untrapped set UT, the minimal dominant set (weak top-cycle) MD. Let $\mathbf{sp}_{(k)}$, $\mathbf{ss}_{(k)}$, $\mathbf{p}_{(k)}$ and $\mathbf{s}_{(k)}$ denote characteristic vectors for classes of k-stable alternatives $SP_{(k)}$, classes of k-stable sets $SP_{(k)}$, and their sums $P_{(k)}=SP_{(1)}+SP_{(2)}+\ldots+SP_{(k)}$, $S_{(k)}=SS_{(1)}+SS_{(2)}+\ldots+SS_{(k)}$, respectively. Let **a** denote a characteristic vector of a universal set A. $\varepsilon$ denotes the relation of identity, which is represented by the matrix $\mathbf{E}=[\delta_{ij}]$. $d=d(\rho)$ is a $\rho$-diameter of A. Let **M**, **T**, **U** denote Boolean matrices representing relations $\mu$, $\tau$ and $\upsilon = \mu \cup \tau \cup \varepsilon$ on A. Finally, let $\mathbf{M}_{(k)} = \sum_{i=1}^{k} \mathbf{M}^i + \mathbf{E}$ and $\mathbf{U}_{(k)} = \sum_{i=1}^{k} \mathbf{U}^i$.

Then

1) $\mathbf{cw} = \overline{\overline{(\mathbf{M}+\mathbf{E}) \cdot \mathbf{a}}}$,

$\mathbf{cr} = \overline{\overline{(\mathbf{M}+\mathbf{T}+\mathbf{E}) \cdot \mathbf{a}}} = \overline{\overline{\mathbf{U}} \cdot \mathbf{a}} = \overline{\mathbf{M}^{tr} \cdot \mathbf{a}}$,

$\mathbf{uc}^I = \overline{\overline{(\mathbf{M} \cdot \mathbf{M}+\mathbf{M}+\mathbf{T}+\mathbf{E}) \cdot \mathbf{a}}} = \overline{\overline{(\mathbf{M} \cdot \mathbf{M}+\mathbf{U}) \cdot \mathbf{a}}}$,

$\mathbf{uc}^I_m = \overline{\overline{(\mathbf{M} \cdot \mathbf{M}+\mathbf{M}+\mathbf{E}) \cdot \mathbf{a}}}$,

$\mathbf{uc}^{II} = \overline{\overline{(\mathbf{M} \cdot \mathbf{T}+\mathbf{M} \cdot \mathbf{M}+\mathbf{M}+\mathbf{T}+\mathbf{E}) \cdot \mathbf{a}}} = \overline{\overline{(\mathbf{M} \cdot \mathbf{U}+\mathbf{U}) \cdot \mathbf{a}}}$,

$\mathbf{uc}^{II}_m = \overline{\overline{(\mathbf{M} \cdot \mathbf{T}+\mathbf{M} \cdot \mathbf{M}+\mathbf{M}+\mathbf{E}) \cdot \mathbf{a}}} = \overline{\overline{(\mathbf{M} \cdot \mathbf{U}+\mathbf{E}) \cdot \mathbf{a}}}$,

$$\mathbf{uc}^{III} = \overline{\overline{(\mathbf{T} \cdot \mathbf{M} + \mathbf{M} \cdot \mathbf{M} + \mathbf{M} + \mathbf{T} + \mathbf{E}) \cdot \mathbf{a}}} = \overline{\overline{(\mathbf{U} \cdot \mathbf{M} + \mathbf{U}) \cdot \mathbf{a}}},$$

$$\mathbf{uc}^{III}{}_m = \overline{\overline{(\mathbf{T} \cdot \mathbf{M} + \mathbf{M} \cdot \mathbf{M} + \mathbf{M} + \mathbf{E}) \cdot \mathbf{a}}} = \overline{\overline{(\mathbf{U} \cdot \mathbf{M} + \mathbf{E}) \cdot \mathbf{a}}},$$

$$\mathbf{uc}^{IV} = \overline{\overline{(\mathbf{T} \cdot \mathbf{M} + \mathbf{M} \cdot \mathbf{T} + \mathbf{M} \cdot \mathbf{M} + \mathbf{M} + \mathbf{T} + \mathbf{E}) \cdot \mathbf{a}}} = \overline{\overline{(\mathbf{U} \cdot \mathbf{M} + \mathbf{M} \cdot \mathbf{U} + \mathbf{U}) \cdot \mathbf{a}}},$$

$$\mathbf{uc}^{IV}{}_m = \overline{\overline{(\mathbf{T} \cdot \mathbf{M} + \mathbf{M} \cdot \mathbf{T} + \mathbf{M} \cdot \mathbf{M} + \mathbf{M} + \mathbf{E}) \cdot \mathbf{a}}} = \overline{\overline{(\mathbf{U} \cdot \mathbf{M} + \mathbf{M} \cdot \mathbf{U} + \mathbf{E}) \cdot \mathbf{a}}},$$

$$\mathbf{uc}^{V} = \overline{\overline{(\mathbf{T} \cdot \mathbf{T} + \mathbf{T} \cdot \mathbf{M} + \mathbf{M} \cdot \mathbf{T} + \mathbf{M} \cdot \mathbf{M} + \mathbf{M} + \mathbf{T} + \mathbf{E}) \cdot \mathbf{a}}} = \overline{\overline{\mathbf{U} \cdot \mathbf{U} \cdot \mathbf{a}}},$$

$$\mathbf{uc}^{V}{}_m = \overline{\overline{(\mathbf{T} \cdot \mathbf{T} + \mathbf{T} \cdot \mathbf{M} + \mathbf{M} \cdot \mathbf{T} + \mathbf{M} \cdot \mathbf{M} + \mathbf{M} + \mathbf{E}) \cdot \mathbf{a}}} = \overline{\overline{(\mathbf{T} \cdot \mathbf{T} + \mathbf{U} \cdot \mathbf{M} + \mathbf{M} \cdot \mathbf{U} + \mathbf{E}) \cdot \mathbf{a}}},$$

$$\mathbf{ucp} = \overline{\overline{(\mathbf{M} \cdot \mathbf{T} \cdot \mathbf{M} + \mathbf{M} \cdot \mathbf{M} \cdot \mathbf{M} + \mathbf{T} \cdot \mathbf{M} + \mathbf{M} \cdot \mathbf{T} + \mathbf{M} \cdot \mathbf{M} + \mathbf{M} + \mathbf{T} + \mathbf{E}) \cdot \mathbf{a}}} =$$

$$= \overline{\overline{(\mathbf{M} \cdot \mathbf{U} \cdot \mathbf{M} + \mathbf{U} \cdot \mathbf{M} + \mathbf{M} \cdot \mathbf{U} + \mathbf{U}) \cdot \mathbf{a}}},$$

$$\mathbf{mws}^{II} = (\mathbf{M} + \mathbf{E}) \cdot \overline{\overline{(\mathbf{T} \cdot \mathbf{M} + \mathbf{M} \cdot \mathbf{M} + \mathbf{M} + \mathbf{T} + \mathbf{E}) \cdot \mathbf{a}}} = (\mathbf{M} + \mathbf{E}) \cdot \overline{\overline{(\mathbf{U} \cdot \mathbf{M} + \mathbf{U}) \cdot \mathbf{a}}},$$

$$\mathbf{mws}^{III} = \overline{\overline{(\mathbf{M} \cdot \mathbf{U} + \mathbf{U}) \cdot \mathbf{a}}} + \mathrm{diag}((\mathbf{M} + \mathbf{T}) \cdot \overline{\overline{(\mathbf{M} \cdot \mathbf{U} + \mathbf{E}) \cdot \mathbf{M}}}),$$

$$\mathbf{mu} = \overline{\overline{(\mathbf{M}_{(d)} + \overline{\mathbf{M}}_{(d)}^{tr}) \cdot \mathbf{a}}}, \ d = d(\mu): (\mathbf{M}_{(d)} \neq \mathbf{M}_{(d-1)}) \ \& \ (\mathbf{M}_{(d)} = \mathbf{M}_{(d+1)}),$$

$$\mathbf{ut} = \overline{\overline{(\mathbf{M}_{(d)} + \mathbf{T}) \cdot \mathbf{a}}}, \ d = d(\mu): (\mathbf{M}_{(d)} \neq \mathbf{M}_{(d-1)}) \ \& \ (\mathbf{M}_{(d)} = \mathbf{M}_{(d+1)}),$$

$$\mathbf{md} = \overline{\overline{\mathbf{U}_{(d)} \cdot \mathbf{a}}}, \ d = d(\upsilon): (\mathbf{U}_{(d)} \neq \mathbf{U}_{(d-1)}) \ \& \ (\mathbf{U}_{(d)} = \mathbf{U}_{(d+1)}).$$

2) If $\mu$ is a tournament, then $\mathbf{T} = \mathbf{O}$, $\mathbf{U} = \mathbf{M} + \mathbf{E}$, $\mathbf{M}_{(k)} = \mathbf{U}_{(k)} = \mathbf{U}^k$ and

$$\mathbf{p}_{(k)} = \overline{\overline{\mathbf{M}_{(k)} \cdot \mathbf{a}}} = \overline{\overline{\mathbf{U}^k \cdot \mathbf{a}}},$$

$$\mathbf{sp}_{(k)} = \overline{\overline{\mathbf{p}_{(k)} + \mathbf{p}_{(k-1)}}} = \overline{\overline{\mathbf{M}_{(k)} \cdot \mathbf{a}}} + \overline{\overline{\mathbf{M}_{(k-1)} \cdot \mathbf{a}}} = \overline{\overline{\mathbf{U}^k \cdot \mathbf{a}}} + \overline{\overline{\mathbf{U}^{k-1} \cdot \mathbf{a}}},$$

$$\mathbf{cw} = \mathbf{p}_{(1)} = \mathbf{sp}_{(1)} = \overline{\overline{(\mathbf{M} + \mathbf{E}) \cdot \mathbf{a}}} = \overline{\overline{\mathbf{U} \cdot \mathbf{a}}},$$

$$\mathbf{uc} = \mathbf{p}_{(2)} = \overline{\overline{(\mathbf{M}^2 + \mathbf{M} + \mathbf{E}) \cdot \mathbf{a}}} = \overline{\overline{\mathbf{U}^2 \cdot \mathbf{a}}},$$

$$\mathbf{ucp} = \mathbf{p}_{(3)} = \overline{\overline{(\mathbf{M}^3 + \mathbf{M}^2 + \mathbf{M} + \mathbf{E}) \cdot \mathbf{a}}} = \overline{\overline{\mathbf{U}^3 \cdot \mathbf{a}}},$$

$$\mathbf{mu} = \mathbf{ut} = \mathbf{md} = \mathbf{p}_{(m)} = \overline{\overline{(\mathbf{M}_{(m)} + \mathbf{E}) \cdot \mathbf{a}}} = \overline{\overline{\mathbf{U}^m \cdot \mathbf{a}}}, \ m: \mathbf{p}_{(m-1)} \neq \mathbf{p}_{(m)} \ \& \ \mathbf{p}_{(m)} = \mathbf{p}_{(m+1)},$$

$$s_{(1)}=ss_{(1)}=mws=(M+E)\cdot p_{(2)}=U\cdot \overline{\overline{U^2 \cdot a}},$$

$$s_{(k)}=s_{(k-1)}+mws^{III}(\upsilon_{(k)})=s_{(k-1)}+\overline{\overline{(\tilde{M}\cdot \tilde{U}+\tilde{U})\cdot a}}+diag(\overline{\overline{(\tilde{M}+\tilde{T})\cdot \overline{\overline{(\tilde{M}\cdot \tilde{U}+E)\cdot \tilde{M}}}}}),$$

$$ss_{(k)}=\overline{\overline{\bar{s}_{(k)}+s_{(k-1)}}},$$

where $\tilde{U}=M_{(k)}$, $\tilde{M}=\overline{\overline{(M_{(k)}^{tr}+\overline{M}_{(k)})}}$.

**Appendix**

**Proof of Lemma 3.** Suppose $i \in B$, B is a minimal weakly stable set (third version), $B \subseteq MWS^{III}$. Then $B\setminus\{i\}$ is not weakly stable $\Leftrightarrow \exists j: (B\setminus\{i\}) \subseteq L(j)$. At the same time if $j \notin B$ then $B \cap (D(j) \cup H(j)) \neq \emptyset$. Consequently either $j=i$ or $i \in D(j) \cup H(j)$ holds, that is $i \in D(j) \cup H(j) \cup \{j\}$. A condition $i \in D(j) \cup H(j) \cup \{j\}$ is equivalent to $j \in L(i) \cup H(i) \cup \{i\}$.

A condition $(B\setminus\{i\}) \subseteq L(j)$ is equivalent to $B \subseteq L(j) \cup \{i\}$. Since by assumption B is a weakly stable set and $B \subseteq L(j) \cup \{i\}$ then $L(j) \cup \{i\}$ must be a weakly stable set as well (monotonicity of weak stability). Consequently, if B is a minimal weakly stable set and $i \in B$ then it is necessary that $\exists j$: $j \in L(i) \cup H(i) \cup \{i\}$ & $L(j) \cup \{i\}$ is a weakly stable set.

Let us prove that this condition is sufficient for the existence of a minimal weakly stable set B such that $i \in B$. Suppose $\exists j$: $j \in L(i) \cup H(i) \cup \{i\}$ & $L(j) \cup \{i\}$ is a weakly stable set. If $L(j) \cup \{i\}$ is minimal then $B=L(j) \cup \{i\}$. If it is not then $\exists C$: $C \subset L(j) \cup \{i\}$ and C is a minimal weakly stable set. By definition $L(j)$ is not a weakly stable set. Since C is weakly stable, C is not a subset of $L(j)$ (monotonicity of weak stability). But $C \subset L(i) \cup \{i\}$, therefore $i \in C$ and $B=C$.

Thus, i belongs to a minimal weakly stable set iff $\exists j$: $j \in L(i) \cup H(i) \cup \{i\}$ and $L(j) \cup \{i\}$ is a weakly stable set.

$L(i) \cup \{i\}$ is not a weakly stable set $\Leftrightarrow \exists k$: $(L(i) \cup \{i\}) \subseteq L(k) \Leftrightarrow k\mu i$ & $L(i) \subseteq L(k) \Leftrightarrow i \notin UC^{II}$. Therefore, $L(i) \cup \{i\}$ is a weakly stable set iff i is uncovered according to the second version of the covering relation, $i \in UC^{II}$.

Suppose ∃j: 1) j∈L(i)∪H(i) & 2) L(j)∪{i} is not a weakly stable set. Then (2) ⇔ ∃k: (L(j)∪{i})⊆L(k). Then (L(j)∪{i})⊆L(k) ⇔ L(j)⊆L(k) & {i}⊆L(k). Then {i}⊆L(k) ⇔ k∈D(i). Since by definition D(i)∩(L(i)∪H(i))=∅ for any i∈A, (j∈L(i)∪H(i) & k∈D(i)) ⇒ k≠j. Then (k≠j & L(j)⊆L(k)) ⇔ j is covered by k according to modified second version of the covering relation. Consequently, ∃j: 1) j∈L(i)∪H(i) & 2) L(j)∪{i} is not a weakly stable set ⇔ ∃j, k: 1) j∈L(i)∪H(i) & 2) k∈D(i) & 3) j is covered by k according to modified second version of the covering relation. Therefore a set L(j)∪{i}: j∈L(i)∪H(i) is weakly stable iff j is not covered by any alternative from the upper contour set of i according to modified second version of the covering relation.

Therefore ∃j: 1) j∈L(i)∪H(i)∪{i} and 2) L(j)∪{i} is a weakly stable set ⇔ either i∈UC$^{II}$, or ∃j: 1) j∈L(i)∪H(i) & 2) j is not covered by any alternative from the upper contour set of i according to modified second version of the covering relation.

As a result, i belongs to a minimal weakly stable set (third version) MWS$^{III}$ iff either x is uncovered according to the second definition of the covering relation, or some alternative from the lower contour set of i or from the horizon of i is is not covered by any alternative from the upper contour set of i according to modified second version of the covering relation.